%% file: transmit_correlation_diversity_final_v1.tex
\documentclass[12pt,letterpaper,final,peerreviewca]{IEEEtran}
\usepackage{amsmath,amssymb,dsfont,stfloats,amsthm,color,subfigure,cite}
\usepackage{hyperref}
\usepackage[pdftex]{graphicx}

\newtheorem{thm}{Theorem}

\newtheorem{lem}{Lemma}
\newtheorem{corol}{Corollary}

\newtheorem{defi}{Definition}

\theoremstyle{definition}

\theoremstyle{remark}
\newtheorem{remark}{Remark}

\theoremstyle{ex}
\newtheorem{ex}{Example}

\def\argmax{\mathop{\rm argmax}}

\def\pa{{\pmb a}}\def\pd{{\pmb d}}\def\ph{{\pmb h}}\def\pw{{\pmb w}}\def\px{{\pmb x}}\def\py{{\pmb y}}\def\pz{{\pmb z}}

\def\pA{{\pmb A}}\def\pB{{\pmb B}}\def\pH{{\pmb H}}\def\pI{{\pmb I}}\def\pP{{\pmb P}}\def\pR{{\pmb R}}\def\pS{{\pmb S}}\def\pU{{\pmb U}}\def\pW{{\pmb W}}\def\pX{{\pmb X}}\def\pY{{\pmb Y}}\def\pZ{{\pmb Z}}\def\p0{{\pmb 0}}

\def\ct{\mathsf{H}}

\def\snr{{\small\textsf{SNR}}}

\input{macros}

\IEEEoverridecommandlockouts

\begin{document}

\title{Fundamental Limits of Correlated Fading MIMO Broadcast Channels: Benefits of Transmit Correlation Diversity}

\author{Junyoung Nam
\thanks{This work was supported by the ICT R\&D program of MSIP/IITP [14-000-04-001]. The material in this paper was presented in part  at the IEEE International Symposium on Information Theory (ISIT),  Jun./Jul. 2014.}
\thanks{J. Nam is with the Wireless Communications Division, Electronics and Telecommunications Research Institute (ETRI), Daejeon, Korea. (e-mail: jynam@etri.re.kr). }
}

\maketitle

\begin{abstract}

We investigate asymptotic capacity limits of the Gaussian MIMO broadcast channel (BC) with spatially correlated fading to understand when and how much transmit correlation helps the capacity. By imposing a structure on channel covariances (equivalently, transmit correlations at the transmitter side) of users, also referred to as \emph{transmit correlation diversity}, the impact of transmit correlation on the power gain of MIMO BCs is characterized  in several regimes of system parameters, with a particular interest in the large-scale array (or massive MIMO) regime. Taking the cost for downlink training into account, we provide asymptotic capacity bounds of multiuser MIMO downlink systems to see how transmit correlation diversity affects the system multiplexing gain. We make use of the notion of joint spatial division and multiplexing (JSDM) to derive the capacity bounds. It is advocated in this paper that transmit correlation diversity may be of use to significantly increase multiplexing gain as well as power gain in multiuser MIMO systems. In particular, the new type of diversity in wireless communications is shown to improve the system multiplexing gain up to by a factor of the number of degrees of such diversity. Finally, performance limits of conventional large-scale MIMO systems not exploiting transmit correlation are also characterized.
\end{abstract}


\section{Introduction}
\label{sec:intro}

Channel fading had been considered as a harmful source to combat with transmit or receive diversity, but since \cite{Fos96,Tel99} independent fading in multiple-antenna channels has been the useful fountain for increasing the degrees of freedom available for wireless communications. As a matter of fact, in many scattering and antenna configuration scenarios the channel coefficients observed at different antennas are correlated. This is generally referred to as spatially correlated fading.
In particular, we refer to transmit (or receive) correlation to indicate that the correlation between the channel coefficients of the transmit (or receive) antennas. 




Spatially correlated fading multiple-input multiple-output (MIMO) channels have been well characterized for a variety of transmit correlation models \cite{Shi00,Chu02,Ver05,Tul05}. Traditionally, transmit correlation has been considered to be a detrimental source  (e.g., as a power offset at high signal-to-noise ratio (SNR) \cite{Loz05}). Some exceptional cases where transmit correlation helps capacity are when the capacity-achieving input covariance is non-isotropic and SNR is sufficiently {low} \cite{Ver05,Tul05} and when channel state information (CSI) is not available at all \cite{Jaf05}. 
The impact of transmit correlation on the ergodic capacity is much less known in the multiuser context, albeit the capacity region of the Gaussian MIMO BC with perfect CSI at both transmitter and receivers is fully understood \cite{Wei06} irrespectively of transmit correlation. The work of \cite{Aln09} extended the sum-rate scaling result of \cite{Sha05} to the special case where all users have a common channel covariance, which concludes that transmit correlation has a fairly detrimental impact on the sum capacity of MIMO BC, in line with the traditional view.

A different line of thought is that transmit correlation can be in fact advantageous from some other perspectives (e.g., CSI feedback overhead, scheduling, and codebook design \cite{Ham08, Tri08, Cle08}) of multiuser MIMO (MU-MIMO) communications, since there exist \emph{diverse} transmit correlations across multiple users in realistic wireless channels. Basically, different transmit correlations indicate different ``large-scale" (or ``long-term") channel directions of users that depend on the scattering geometry so that the diversity of transmit correlations can be leveraged in the multiuser communication framework. For such effect, we coin the term \emph{transmit correlation diversity}.
Imposing a structure on transmit correlations to fully exploit this diversity, the authors in \cite{Nam13a,Nam13b} characterized the asymptotic capacity behavior of a family of correlated fading MIMO BCs in the large number of users regime. Interestingly, it turned out that the sum capacity of ``spatially well-colored" MIMO BCs may be fairly larger than that of the ``spatially white" MIMO BC in terms of power gain (the parallel shift of capacity versus SNR curves, also known as power offset). However, it was not fully understood why we could do better than the independent and identically distributed (i.i.d.) Rayleigh fading case in this regime. In addition, we do not know whether transmit correlation diversity can promise any capacity gain over the independent fading case in other regimes of interest as well. To this end, we need to investigate the impact of transmit correlation on power gain in different regimes. A main goal of this work is to address these essential questions.


Taking the overhead for downlink training into consideration, a key fundamental limit on the sum rate of the i.i.d. Rayleigh block-fading MIMO BC consisting of a transmitter with $M$ antennas and $K$ receivers (users) with a single antenna each immediately follows from the work of Zheng and Tse \cite[Sec. V]{Zhe02} (see also \cite{Has03}) by allowing cooperation among users. Namely, the high-SNR capacity of the resulting pilot-aided systems is limited by 
\begin{align}  \label{eq:intro-1}
  M_\text{iid}^*(1-M_\text{iid}^*/T_c)\log \snr+O(1)
\end{align}
where $T_c$ is the coherence time interval\footnote{The unit of coherence time interval can be represented as the number of transmit symbols because each training signal is transmitted over a symbol interval, which  in turn corresponds to a channel use.} and $M_\text{iid}^* = \min\{M,K,\lfloor T_c/2\rfloor\}.$
For typical cellular downlink systems with $M$ small, where $\min\{M,K\} \ll T_c$, the factor $T_c/2$ does not significantly affect the system performance. However, in the large-scale array regime with $M > T_c$, to which great attention has been paid in practice since \cite{Mar10}, this factor is shown to have a critical impact on the system performance. To be specific, no matter how large both $M$ and $K$ are, multiplexing gain is fundamentally \emph{saturated} by $T_c/4$ according to (\ref{eq:intro-1}) when $\min(M,K)\ge T_c/2$, which was also observed in \cite{Huh12}. Therefore, the system is not \emph{scalable} in $\min(M,K)$ and the user throughput vanishes as $O(\frac{1}{K})$. This limit holds also in time division duplexing (TDD) systems as in \cite{Mar10}. As a result, for both $M$ and $K$ large, which is the case of large-scale MIMO, 
the coherence time $T_c$ 
becomes a serious limiting factor in the performance of MIMO wireless communications. 
It is another main goal of this work to show that this is not necessarily the case in spatially correlated fading channels.

Throughout this paper, we restrict our attention to an optimistic condition to intuitively expose potential gains of transmit correlation diversity and to provide some new insights into capacity limits of correlated fading BCs. The ideal condition is called the \emph{tall unitary structure} of channel covariances of users for which users in group $g$ have the same channel covariance of rank $r\le M$ for all $g$ and the eigenspaces of all groups are orthogonal to each other. The tall unitary structure was introduced by the authors of \cite{Nam12} to achieve a large-scale MIMO gain in realistic frequency division duplexing (FDD) systems. The resulting JSDM strategy was subsequently extended to present the feasibility of the tall unitary structure and 3-D beamforming for large $M$ \cite{Adh13}, the impact of transmit correlation to the capacity for large $K$ \cite{Nam13a}, a per-group opportunistic beamforming scheme with probabilistic scheduling \cite{Nam13b}, and the suitability of JSDM for millimeter wave (mm-Wave) channels \cite{Adh13b}. Assuming the ideal condition, the number of degrees of transmit correlation diversity is given by an integer $G = \lfloor{M}/{r}\rfloor$. It was shown by\cite{Nam12,Adh13} that JSDM exploiting transmit correlation diversity can reduce pilot overhead (both in FDD and TDD downlinks) and CSI feedback overhead \emph{at least} by a factor of $G$, since the effective channel dimension is reduced by the same factor $G$. However, the following question has not been addressed: 
{How does this pilot saving affect the multiplexing gain of realistic (pilot-aided) MU-MIMO systems taking downlink training into account? This is closely related to the question as to whether we can mitigate or even eliminate the degree-of-freedom saturation effect imposed by (\ref{eq:intro-1}), where the system multiplexing gain is limited regardless of $M$ and $K$. 

In order to explore the impact of transmit correlation on the system multiplexing gain, we reconsider the Zheng-Tse bound (\ref{eq:intro-1}) in the correlated fading BC context. The resulting capacity result can be summarized as follows:
\begin{itemize}
\item Assuming the tall unitary structure with $G$ degrees of transmit correlation diversity, the high-SNR capacity of pilot-aided MU-MIMO systems is upper-bounded by
\begin{align} \label{eq:intro-4}
  M^*\left(1-\frac{M^*}{T_cG}\right) \log\snr + O(1) 
\end{align}
where $M^* = \min\{M,K,\lfloor \frac{T_cG}{2}\rfloor\}$. 
\end{itemize}

\begin{figure} 
\center \includegraphics[scale=.8]{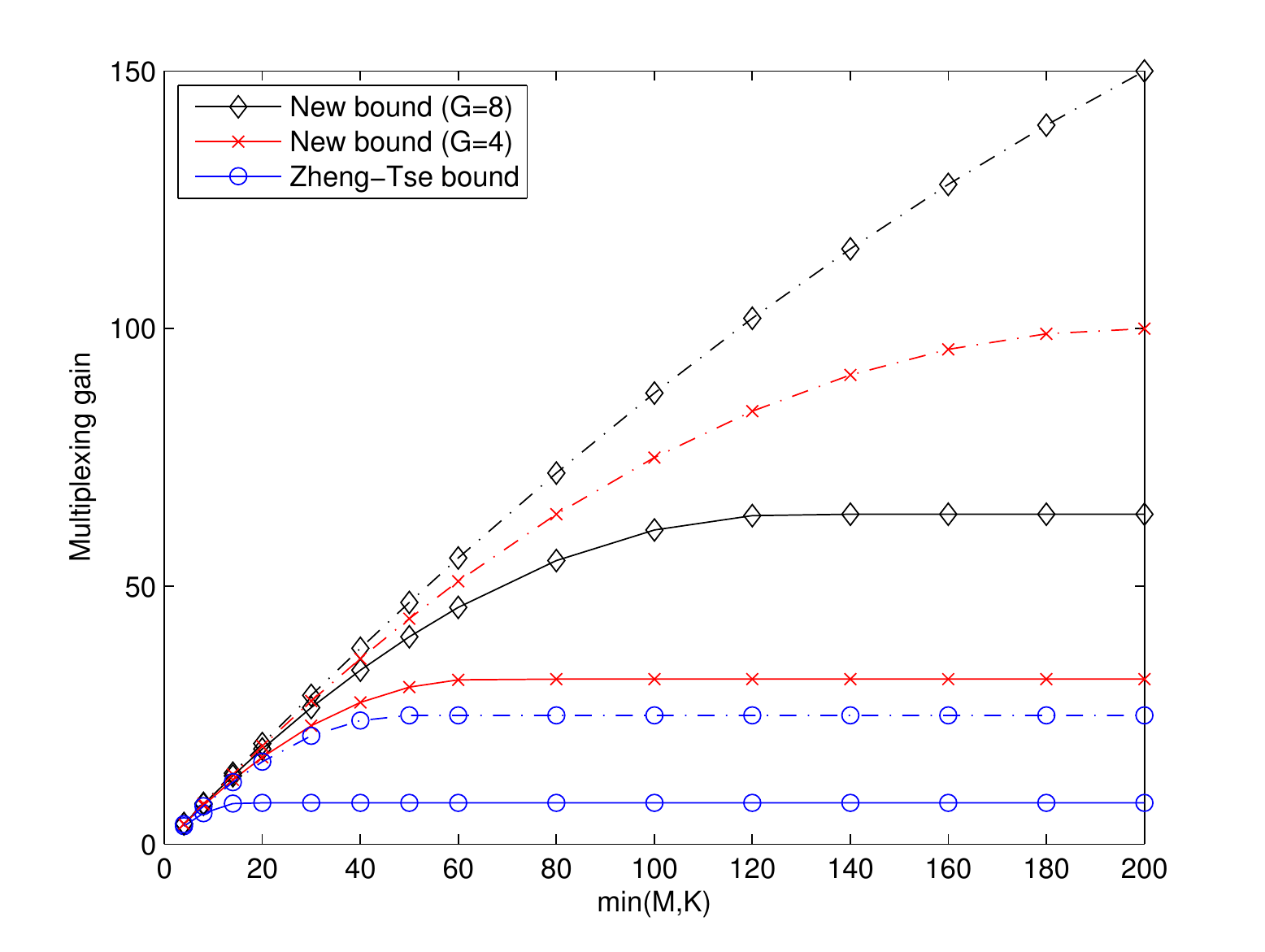} 
\caption{Impact of the degrees of transmit correlation diversity ($G$) on multiplexing gain for different numbers of $\min(M,K)$ in pilot-aided systems, where the solid lines indicate $T_c=32$ and the dash-dotted lines indicate $T_c=100$ (see Example 1 for more details).} \label{fig-1}
\end{figure}

This important result advocates that multiplexing gain can continue growing as $M$ and $K$ increase independently of $T_c$, provided that $G\ge 2\min(M,K)/T_c$, i.e., transmit correlations of users are sufficiently high and well structured.  
This further indicates that transmit correlation diversity can be leveraged to significantly increase the multiplexing gain of MU-MIMO systems, as well as the power gain. 
To show how large the potential gain of transmit correlation diversity could be, Fig. 1 depicts the gap between the multiplexing gains of (\ref{eq:intro-1}) and (\ref{eq:intro-4}). 

Finally, performance limits of conventional large-scale MIMO systems using the uplink/downlink channel reciprocity but not exploiting transmit correlation diversity are further investigated by a practical consideration on the cost of dedicated pilot for coherent detection. This explains why the use of transmit correlation diversity is particularly essential in those systems.

The remainder of this paper is organized as follows. Section \ref{sec:SD} describes the MU-MIMO downlink system model of interest and Section \ref{sec:TCD} briefly reviews a key result of JSDM with the notion of transmit correlation diversity. In Section \ref{sec:IT}, we study the impact of transmit correlation to the power gain of MIMO BCs in several regimes of system parameters. Asymptotic capacity behaviors with focus on the impact of transmit correlation to the multiplexing gain of pilot-aided MU-MIMO systems are presented in Section \ref{sec:FL}. Section \ref{sec:PL} presents performance limits of large-scale MIMO systems. We conclude this work in Section \ref{sec:CR}.

{\em Notation}: 
$\pA^\ct$, $\|\pA\|_F$, and $\lambda_i(\pA)$ denote the Hermitian transpose, the Frobenius norm, the $i$th eigenvalue (in descending order) of matrix $\pA$. $\trace(\pA)$ and $|\pA|$ denote the trace and the determinant of a square matrix $\pA$. $\pI_n$ denotes the $n \times n$ identity matrix.  $\|\pa\|$ denotes the $\ell_2$ norm of vector $\pa$. We also use  $\px \sim \mathcal{CN}(\p0;\Sigmam)$ to indicates that $\px$ is a zero-mean complex circularly-symmetric Gaussian vector with covariance $\Sigmam$. Finally, $\mathbb{Z^+}$ denotes the set of positive integers.

\section{System Description}
\label{sec:SD}

In this paper, we consider a family of spatially correlated Rayleigh fading channels obeying the well-known Kronecker model \cite{Shi00,Chu02} (or separable correlation model)
\begin{align} \label{eq:SM-1}
   \pH={\pR_\mathrm{T}}^{\frac{1}{2}} \Wm {\pR_\mathrm{R}}^{\frac{1}{2}}
\end{align}
where the elements of $\Wm$ are i.i.d. $\sim \mathcal{CN}(0,1)$, and
$\pR_\mathrm{T}$ and $\pR_\mathrm{R}$ denote the deterministic transmit and
receive correlation matrices, respectively, assuming the wide-sense stationarity of the channels. The random matrix $\pH$ follows the frequency-flat block-fading model for which it remains constant during the coherence time interval of $T_c$ but changes independently every interval. Most of our results in this paper remain valid in the more general unitary-independent-unitary model (for which see \cite{Tul05}), since the elements of $\pW$ are allowed to be independent \emph{nonidentically} distributed to apply some well-known results of random matrix theory to be used in this paper. 

Consider a MIMO BC (downlink) with $M$ transmit antennas and $K$ users equipped with a single antenna
each. Since users have no receive correlation in this case,
we let $\pR\triangleq\pR_\mathrm{T}$ of rank $r$ for notational simplicity. Also, let $\pR$ normalized as $\trace(\pR)=M$ without loss of generality for all users. We assume that $r$ non-zero eigenvalues of $\pR$ are uniformly bounded, i.e., for any $r \in \mathbb{Z^+}$, there exists a finite positive constant $\zeta$ such that 
\begin{align} \label{eq:SM-1b}
   \frac{\lambda_1(\pR)}{\lambda_{r}(\pR)}\le \zeta < \infty
\end{align}
where $\lambda_i(\pR)$ is the $i$th eigenvalue of $\pR$ in descending order.
This assumption may seem unrealistic since $\pR$ is generally of full algebraic rank even if eigenvalues except dominant ones decay quickly. However, it is quite reasonable at least in the large number of antennas regime with the antenna configuration of uniform linear array (ULA), for which it was shown in \cite{Adh13} that non-zero eigenvalues of $\pR$ can be accurately approximated by a set of samples $\{S([m/M]): m=0,\cdots,M-1\}$ (with $[x]$ being $x$ modulo the interval $[-1/2,1/2]$) which has support of length $\rho\le 1$ on such an interval. Here $S(\cdot)$ is the eigenvalue spectrum of $\pR$, which will be discussed later in Sec. \ref{sec:IT-B}. 
This implies that non-dominant eigenvalues go to zero when $M$ is sufficiently large.

In this paper, we will sometimes make use of the specific and more realistic one-ring model for $\pR$, which corresponds to the typical cellular downlink case where the basestation (BS) is elevated and free of local scatterers, and the user terminals are placed at ground level and are surrounded by local scatterers. For this case, the channel in the form of (\ref{eq:SM-1}) may reduce to the one-ring model. For the one-ring model, a user located at azimuth angle $\theta$ and distance ${\sf s}$ is surrounded by a ring of scatterers of radius  ${\sf r}$ such that angular spread (AS) $\Delta \approx \arctan({\sf r}/{\sf s})$. Assuming the ULA with a uniform distribution of the received power from planar waves impinging on the BS array, the correlation coefficient between BS antennas $1 \leq p, q \leq M$ is given by
\begin{align} \label{eq:SM-4}
   [\pR]_{p,q} =\frac{1}{2\Delta}  \int_{-\Delta}^{\Delta}e^{j2\pi D(p-q)\sin(\alpha+\theta)}d\alpha
\end{align}
where $D$ is the normalized distance between antenna elements by the wavelength.

By using the Karhunen-Loeve transform, the channel vector of a user can be expressed as
\begin{align} \label{eq:SM-2}
   \ph={\pU}{\boldsymbol \Lambda}^{\frac{1}{2}} \wv
\end{align}
where 
${\boldsymbol \Lambda}$ is an $r\times r$ diagonal matrix whose elements are the non-zero eigenvalues
of $\pR$, $\pU \in \mathbb{C}^{M\times r}$ is a tall unitary matrix whose columns are the eigenvectors of $\pR$ corresponding to the non-zero eigenvalues, i.e., $\pR=\pU\Lambdam\pU^\ct$, and $\wv \in \mathbb{C}^{r \times 1} \sim\mathcal{CN} (\p0, \pI)$. 

Let $\underline{\pH}$ denote the $M\times K$ system channel matrix given by
stacking the $K$ users channel vectors $\ph$ by columns. The signal vector received by the users is given by
\begin{align} \label{eq:SM-3}
   \py=\underline{\pH}^\mathsf{H}\Vm\pd +\pz =\underline{\pH}^\mathsf{H}\px +\pz
\end{align}
where $\Vm$ is the $M\times s$ precoding matrix with $s$ the rank of the input covariance
$\boldsymbol{\Sigma}=\mathbb{E}[\px\px^\ct]$ (i,e., the total number of independent data streams),
$\pd$ is the $s$-dimensional transmitted data symbol vector such that the transmit signal vector is given by $\px=\Vm \pd$, and $\pz  \sim\mathcal{CN} (\p0, \pI)$ is the Gaussian noise
at the receivers. The system has the total power constraint such that $\trace(\Sigmam)\le P$, where $P$ implies the total transmit SNR.

We briefly review next the JSDM strategy \cite{Nam12} that was originally introduced to reduce the cost for downlink training and CSI feedback in FDD large-scale MIMO systems by exploiting the fact that some users have similar transmit correlation matrices
and further by imposing a useful structure on transmit correlations of users.
In order to create a useful structure on transmit correlations, for a given user and scatterer geometry, we put together users with similar transmit correlations into a group and then separate multiple groups by {spatial division}, whose ``long-term" subspaces are quasi-orthogonal. 
In general, we have multiple sets of quasi-orthogonal groups, which we call \emph{classes}. Each class $t$ is served with orthogonal time/frequency resources and may have a different number of groups per class, denoted by $G_t$. Therefore, we partition the entire user set, $\mathcal{K}=\{1,2,\cdots,K\}$, into $T$ non-overlapping subsets (classes).

The precoding of JSDM has a two-stage
structure given by  $\Vm = \pB \pP$, where $\pB
\in\mathbb{C}^{M\times b}$ with $b\le M$ is a {\em pre-beamforming} matrix that
depends only on the channel second-order statistics and $\pP
\in\mathbb{C}^{b\times s}$ is a precoding matrix that depends on
the instantaneous realization of $\underline{\pH}^\mathsf{H}\pB$.
We divide $b$ such that $b=\sum_g b_g$, where $b_g$ is an integer not smaller than $s_g$ (the number of independent data streams for group $g$),
and denote by $\pB_g$ the $M\times b_g$ pre-beamforming matrix of group $g$.
Thanks to the above user partitioning, we can consider feeding back only the $G$ diagonal blocks 
\begin{align} \label{eq:SM-10}
   \textsf{\pH}_g \triangleq \Bm_g^\mathsf{H} \Hm_g, \ g=1,\cdots,G
\end{align} 
where $\pH_g$ is the aggregate channel matrix given by stacking the channel vectors of users in group $g$, 
and each group is independently processed by treating signals of other groups as interference due to the quasi-orthogonality of groups within a class. 
In this case, the precoding matrix takes on the block-diagonal form
$\pP=\mathrm{diag}(\pP_1,\cdots,\pP_G)$, where $\pP_g\in \mathbb{C}^{b_g \times s_g}$, yielding the vector broadcast plus interference Gaussian channel
 $\py_g = {\pH}_g^\mathsf{H}\pB_g \pP_g\pd_g +  \sum_{h \neq g}  {\pH}_g^\mathsf{H}\pB_{h} \pP_{h}\pd_{h}   +  \pz_g$ for all $g$. 

\section{Transmit Correlation Diversity}
\label{sec:TCD}

In this section, we introduce the terminology of transmit correlation diversity to better understand the key idea of JSDM. Fig. \ref{fig-0} depicts a simple example which explains the virtual sectorization enabled by exploiting diverse transmit correlations in a three-sector BS with $D=1/2$ in the following steps.  

\begin{enumerate}
\item In the beginning, angular regions (pie slices drawn by AoD\footnote{AoD and angle of arrival (AoA) are generally different in FDD. As AoD is more precise at the transmitter side, we prefer the terminology AoD to AoA.} 
 and AS at the BS) roughly representing long-term eigenspaces are overlapped, i.e., user groups are interfering with each other and there is no noticeable structure. 
\item Put together the red angular regions into class $t=1$ and separate them by multiple pre-beamforming along their respective eigenspaces. By doing so, each group can be viewed as a virtual sector.
\item Do the same thing on the blue regions for class $t=2$.
\item Multiple users within each group (i.e., virtual sector) can be simultaneously served by MU-MIMO precoding.
\end{enumerate}

\begin{figure}
\center \includegraphics[scale=1]{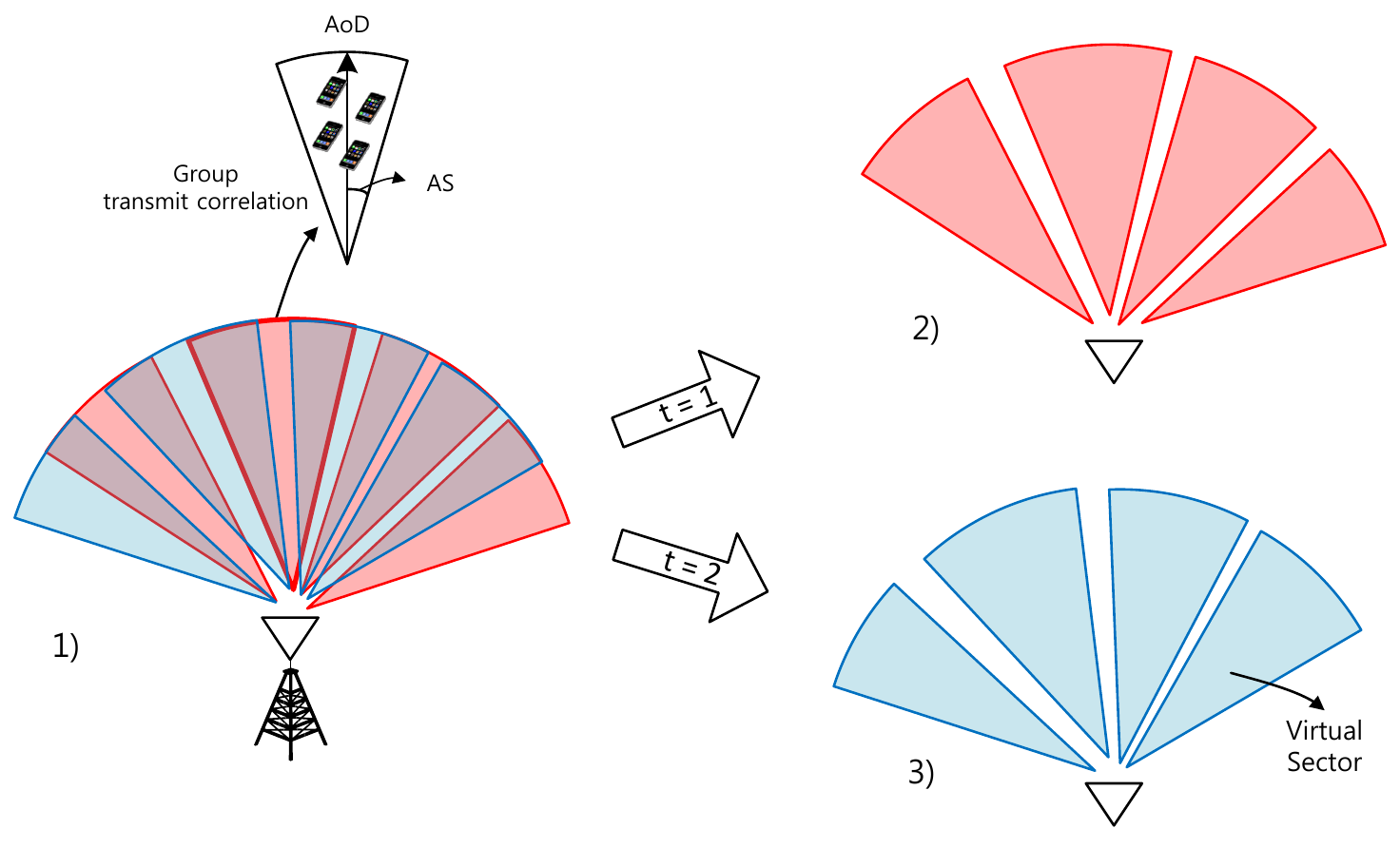} 
\caption{Illustration of virtual sectorization exploiting transmit correlation diversity with $T=2$ and $G=4$.}  \label{fig-0}
\end{figure}

Given the geometric intuition provided by the clustered scattering correlation model (e.g., one-ring model), we define transmit correlation diversity of a multiuser system as follows:

\begin{defi}[Transmit Correlation Diversity] \label{def-3}
A multiuser MIMO downlink system after user partitioning is said to have $G$ degrees of transmit correlation diversity, if the eigenspaces of $G_t$ groups in class $t$ are orthogonal to each other for all $t$ and if $G=\big\lfloor\frac{1}{T}\sum_{t=1}^T G_t\big\rfloor$.
\end{defi}

Throughout this work, we let $T=1$ and assume that $G$ groups are formed in a symmetric manner such that each group has the same number $K'=K/G$ of users and the same rank $r=M/G$ of $\pR_g$ for simplicity, where $G$ divides both $K$ and $M$. It is easy to extend to the general case of multiple classes and asymmetric per-group parameters. In the sequel, we present the ideal structure of transmit correlations at the transmitter and a key result in the prior work \cite{Nam12} for self-containment.

\begin{defi}[Unitary Structure] \label{def-1}
A unitary structure of transmit correlations is obtained if users in group $g$ have a common $\pU_{g}$ for all $g$ and if the $M \times rG$ matrix $\underline{\pU}=[\pU_1,\cdots,\pU_G]$ is {unitary} such that $ \underline{\pU}^\mathsf{H} \underline{\pU}=\underline{\pU} \underline{\pU}^\ct = \Id.$
\end{defi}

For the general case of $rG< M$, the ideal structure is called tall unitary such that $ \underline{\pU}^\mathsf{H} \underline{\pU}= \Id$.
Under the unitary structure, we just let $b_g = r$ and $\Bm = \underline{\pU}$. These choices eliminate interference between $G$ groups and the resulting MIMO BC is given by
\begin{align} \label{eq:SM-5}
   \py_g & = \textsf{\pH}_g^\mathsf{H} \pP_g\pd_g +\pz_g = \Wm_g^\mathsf{H} \Lambdam_g^{1/2} \pP_g\pd_g +\pz_g
\end{align}
where $\Wm_g$ is an $r \times K'$ matrix with i.i.d. elements $\sim \mathcal{CN}(0,1)$, for all $g$, thereby yielding the reduced column dimensionality of the effective channel $\textsf{\pH}_g$. Using (\ref{eq:SM-5}), we arrive at the following simple yet important result.

\begin{thm}[\cite{Nam12}] \label{thm-1}
Under the unitary structure, the ergodic sum capacity of the original MIMO BC
(\ref{eq:SM-3}) with full CSI is equal to that of parallel subchannels (\ref{eq:SM-5}) with reduced CSIT $\textsf{\pH}_g$, given by
\begin{align} \label{eq:MU-0}
 \sum_{g=1}^G \mathbb{E}_{\textsf{\pH}_g} \bigg[\max_{\sum_{g=1}^G\trace(\pS_g)\le P} \; \log  \Big| \Id + \Lambdam_g^{1/2} \Wm_g \Sm_g \Wm_g^\mathsf{H} \Lambdam_g^{1/2}  \Big|\bigg]
\end{align}
where $\Sm_g$ denotes the diagonal $K' \times K'$ input covariance matrix for group $g$ in the dual multiple-access channel (MAC).
\end{thm}

This can be intuitively verified by noticing that the effective channel $\textsf{\pH}_g$ with reduced dimension of $K'\times r$ is unitarily equivalent to the original channel $\pH_g$ of $K'\times M$ under the unitary condition so that we can get \emph{effective channel dimension reduction} without loss of optimality. 
This dimension reduction effect provides significant savings in CSI feedback at least by a factor of $G$. 

\section{Impact of Transmit Correlation to the Power Gain of MIMO BCs}
\label{sec:IT}

It was shown in \cite{Nam13a} that in the large number of users regime, transmit correlation may significantly help the capacity of Gaussian MIMO BCs. One may think that if we fully exploit the ideal unitary structure, we might do better than the independent fading case in other regimes of interest as well. We will show that this holds true in some cases but not in general. To this end, we find out that there is a tradeoff between power loss due to the effective channel dimension reduction and beamforming gain from pre-beamforming across the long-term eigenspace for each user group. In order to understand the tradeoff, we carefully investigate the impact of transmit correlation on the power gain of correlated fading MIMO BCs for different regimes in terms of $r$, $K'$, and $G$. 

In the sequel, we first consider the asymptotic capacity bounds of correlated fading MIMO-BCs in the high-SNR regime, also characterize the high-SNR capacity in the large $M$ regime in a compact form, and then compare these results with those of the independent fading case to see if there exist benefits of transmit correlation to the power gain of the channels.

\subsection{High-SNR Analysis}
\label{sec:IT-A}

For $M$ fixed, we investigate ergodic capacity bounds of MIMO BC at high SNR to capture a rate gap (if any) between the i.i.d. Rayleigh fading channel and the correlated Rayleigh fading channel satisfying the unitary structure. 
Let $\Cc^\text{sum}(a, b)$ denote the asymptotic sum capacity when system parameters $a$ and $b$ are sufficiently large. Since a closed-form characterization of the ergodic sum capacity of MIMO BCs is very little known even in the full CSI case \cite{Gol03}, we rely on some upper and lower bounds on the capacity.
Using Theorem \ref{thm-1} and some well-known results about MIMO BC and random matrix theory in Appendix \ref{app:lem}, we get the following simple bounds on the high-SNR capacity behaviors of correlated fading MIMO BCs.

\begin{thm} \label{thm-2}
Suppose the perfect CSIT on ${\textsf{\pH}}_g$ in (\ref{eq:SM-10}) and the unitary structure on  channel covariances of users in Definition \ref{def-1}. For $r<K'$, the high-SNR capacity of the corresponding MIMO BC with correlated Rayleigh fading behaves like
\begin{align}  \label{eq:IT-13}
  \mathcal{C}^\text{sum}(P) =  M \log\frac{P}{M}+\sum_{g=1}^G\log|\Lambdam_g|+\kappa(K',r)+c_{P,1}+o(1)
\end{align}
where $\kappa(x,y)= yG \big(-\gamma +\sum_{\ell=2}^{x}\frac{1}{\ell}+\frac{x-y}{y}\sum_{\ell=x-y+1}^{x}\frac{1}{\ell} \big)\log e$ with $\gamma$ the Euler-Mascheroni constant, $o(1)$ goes to zero as $P\rightarrow \infty$, and the constant term $c_{P,1}$ is bounded by $$-M\log \frac{K'}{r} \le c_{P,1}\le 0.$$
For $r\ge K'$
\begin{align}  \label{eq:IT-14}
  \mathcal{C}^\text{sum}(P)   &=  K \log\frac{P}{K}+\sum_{g=1}^G \sum_{i=1}^{K'}\log \lambda_{g,i}+\kappa(r,K')+c_{P,2}+o(1)
\end{align} 
where $c_{P,2}$ is bounded by $$\sum_{g=1}^G\sum_{i=1}^{K'}\log\frac{\lambda_{g,r-i+1}}{\lambda_{g,i}} \le c_{P,2}\le 0.$$
\end{thm}

\begin{IEEEproof} 
See Appendix \ref{app:proof-2}.
\end{IEEEproof}

The above result can be generalized to the tall unitary structure for which $rG \le M$ and $M$ in (\ref{eq:IT-13}) and (\ref{eq:IT-14}) is replaced by $rG.$  Notice that the second terms in (\ref{eq:IT-13}) and (\ref{eq:IT-14}) indicate power gain due to multiple pre-beamforming along long-term eigenspaces which we call \emph{eigen-beamforming gain}. 
The bounds in (\ref{eq:IT-13}) of the $r< K'$ case may become arbitrarily loose when $r\ll K'$. This is because the multiuser diversity gain unique in the multiuser framework cannot be expressed in the above formula based on asymptotic point-to-point equivalence. Nevertheless, we will use these bounds in the sequel since the looseness is limited as long as $K'/r$ is not too large. The upper bound in (\ref{eq:IT-13}) is asymptotically tight when the receivers are allowed to cooperate \emph{inside} each group\footnote{Since users in a particular group are often closely located, the partial cooperation within such a group seems more feasible than the full cooperation across all users over the entire BS coverage.}, which we call \emph{partial cooperation} in this work. 
In the case of $r=K'$, (\ref{eq:IT-13}) equals to (\ref{eq:IT-14}) and it is also asymptotically tight with $c_{P,1}=c_{P,2}=0$.

\begin{remark}
An alternative expression of the asymptotic capacity behavior for $r\ge K'$ can be found in (\ref{eq:IT-43}) of Appendix \ref{app:proof-2}, based on the approach in \cite{Shi03,Loz05}\footnote{Although these point-to-point results assume that only the distribution of a channel  is accessible at the transmitter, the difference with the perfect CSIT case that we are assuming vanishes at high SNR when the number of receive antennas is greater than or equal to the total number of transmit antennas (this is the case of the dual MAC in (\ref{eq:MU-0}) when $r\ge K'$).}.
Comparing with the alternative characterization and other previous results \cite{Ver05,Tul05} for the point-to-point MIMO case, we can see that (\ref{eq:IT-14}) in Theorem \ref{thm-2} is more intuitive and insightful. For example, (\ref{eq:IT-14}) will be used in Sec. \ref{sec:IT-D} to show that, for $r \ge K'$, in general we cannot do better than the independent fading case. 
\end{remark}

It immediately follows from \cite{Loz05} and \cite{Cai03} that, for $r\ge K'$ with $G=1$, the capacity of the i.i.d. Rayleigh fading MIMO BC with full CSI behaves like
\begin{align}  \label{eq:IT-50}
  \mathcal{C}^\text{sum}_\text{iid}(P) = M \log\frac{P}{M}+\kappa(M,K)+o(1).
\end{align}

\subsection{Large System Analysis}
\label{sec:IT-B}

We turn our attention to the large number of antennas regime, i.e., the large system analysis. 
In this case, we need the asymptotic behavior of large-dimensional Wishart matrices. To this end, the best known approach is using the Mar\v{c}enko-Pastur law \cite{Mar67}. In this paper, we shall instead use the asymptotic behavior of (\ref{eq:ID-4}) in Appendix \ref{app:lem} to avoid a much more involved definite integral calculation than \cite{Ver99}. 

Let $$\mu=\frac{M}{K}=\frac{r}{K'}$$ and $G$ be fixed such that both $r$ and $K'$ are taken to infinity along with $M$.

\begin{thm} \label{thm-8}
Suppose the perfect CSIT on ${\textsf{\pH}}_g$ and the unitary structure. As $M\rightarrow \infty$, for $\mu < 1$,  the high-SNR capacity of the corresponding correlated fading MIMO BCs scales linearly in $M$ with the ratio
\begin{align}  \label{eq:ID-5}
  \frac{\mathcal{C}^\text{sum}(P,M,r)}{M} = \log \frac{P}{e\mu}+\Big(\frac{1-\mu}{\mu}\Big)\log\frac{1}{1-\mu} +c_{M,1}+o(1)
\end{align}
where the constant $c_{M,1}$ is bounded by $$\log \frac{\mu\lambda_\text{min}}{G} \le c_{M,1}\le 0$$
with $\lambda_\text{min}=\min_g \lambda_{g,r}$.
For $\mu \ge 1$, the high-SNR capacity scales linearly in $K$ with the ratio
\begin{align}  \label{eq:ID-6}
  \frac{\mathcal{C}^\text{sum}(P,M,r)}{K}  &= \log \frac{\mu P}{e}+(\mu-1)\log\frac{\mu}{\mu-1} +c_{M,2}+o(1)
\end{align}
where $c_{M,2}$ is bounded by $$\log \frac{\lambda_\text{min}}{G} \le c_{M,2}\le \log \mu.$$
\end{thm}

\begin{IEEEproof} 
See Appendix \ref{app:proof-8}.
\end{IEEEproof}

When $M=K$ with $\pR_g=\pI_r$ for all $g$, we can easily see that  (\ref{eq:ID-6}) reduces to 
\begin{align} \label{eq:ID-23}
   \frac{\Cc^\text{sum}(P,M)}{M} =\log \frac{P}{e}+o(1)
\end{align}
which equals the well-known ratio of the i.i.d. Rayleigh fading MIMO channel \cite{Fos96}. The asymptotic capacity of the i.i.d. Rayleigh fading MIMO BC is upper-bounded by the point-to-point case (c.f., \cite[Proposition 2]{Loz05})
\begin{align}  
  \frac{\mathcal{C}^\text{sum}_\text{iid}(P,M,r)}{M} &= \log \frac{P}{e\mu}+\Big(\frac{1-\mu}{\mu}\Big)\log\frac{1}{1-\mu} +o(1), \ \text{if } \mu< 1 \label{eq:ID-5b} \\
  \frac{\mathcal{C}^\text{sum}_\text{iid}(P,M,r)}{K}  &= \log \frac{\mu P}{e}+(\mu-1)\log\frac{\mu}{\mu-1} +o(1), \hspace{3mm} \text{if } \mu\ge 1. \label{eq:ID-6b}
\end{align}



In this large $M$ regime, we need to better understand the asymptotic behavior of the logarithm $\log|\pR_g|=\log|\Lambdam_g|$ of the determinant of the transmit correlation matrix $\pR_g$ for group $g$. For any Hermitian positive definite $\pR_g$ and large but finite $M$, we have to rely on some bounds of $\log|\Lambdam_g|$. In this work, we make use of the simple bounds \cite{Bai96} in Lemma \ref{lem-4} of Appendix \ref{app:lem}. Since the upper bound is given by (\ref{eq:ID-20}), it suffices to get a lower bound. Noticing the identity $\log |\pA|=\trace(\log \pA)$ for a Hermitian positive definite matrix $\pA$ and further bounding the lower bound in Lemma \ref{lem-4} through $\trace(\Lambdam_g^2)\le \trace(\Lambdam_g)^2=M^2$, we can easily show that 
\begin{align} \label{eq:ID-28}
  \log \left | \Lambdam_g \right | &\ge \frac{1}{\lambda_\text{min} \tau^2 -\lambda_\text{min}^2 \tau} \bigg\{(\tau^2M-\tau M^2)\log \lambda_\text{min} +(-\lambda_\text{min}^2M +\lambda_\text{min}M^2)\log \tau \bigg\} \nonumber \\
  &= \frac{rG}{\lambda_\text{min}G-\lambda_\text{min}^2(\frac{\lambda_\text{min}-G}{\lambda_\text{min}-rG})} \Bigg\{\bigg(G-\frac{(\lambda_\text{min}-G)rG}{\lambda_\text{min}-rG}\bigg)\log \lambda_\text{min} \nonumber \\
  &\ \ \ \ +\bigg(\frac{(\lambda_\text{min}r-\lambda_\text{min}^2G)(\lambda_\text{min}-G)^2}{(\lambda_\text{min}-rG)^2}\bigg)\log \frac{(\lambda_\text{min}-rG)G}{\lambda_\text{min}-G} \Bigg\}
\end{align}
where $\lambda_\text{min}=\min_g \lambda_{g,r}$ and $\tau=\frac{\lambda_\text{min}M-M^2}{\lambda_\text{min}r-M}$. This lower bound involves only $\lambda_\text{min}$ and the system parameters of interest $r$ and $G$. Therefore, we can obtain from (\ref{eq:ID-28}) a simple lower bound for general $\pR_g$ with $M$ large but finite. It is evident that $\lim_{r\rightarrow \infty}\frac{1}{M}\sum_{g=1}^G\log|\Lambdam_g|  \ge \log\lambda_\text{min}$, which is in line with the lower bound on $c_{M,1}$ in (\ref{eq:ID-5}). 

Assuming the one-ring model in Section \ref{sec:SD} on $\pR$ with the ULA antenna configuration, the transmit correlation matrix can be given in the Hermitian Toeplitz form (\ref{eq:SM-4}) with geometric (large-scale) channel parameters AoD $\theta$, AS $\Delta$, and the normalized antenna spacing $D$. The eigenvalue spectrum $S(\xi)$ of $\pR$ is defined by the Fourier transform of the coefficients $r_k\triangleq[\pR]_{\ell,\ell-k}$, i.e., 
$$S(\xi)= \sum_{k=0}^\infty r_k e^{-j2\pi k\xi}, \ \xi \in [-1/2,1/2]$$
which is a uniformly bounded absolutely integrable function over $\xi$. Then, the limiting behavior of $\frac{1}{r}\log|\pR| $ can be explicitly expressed by using the well-known Szeg\"{o} theorem \cite{Gre84, Gra06} on the asymptotic behavior of eigenvalues of a sequence of Hermitian Toeplitz matrices as follows: 
\begin{align} \label{eq:ID-29}
   \lim_{r\rightarrow \infty}\frac{1}{r}\log|\pR| = \int_{-1/2}^{1/2} \log S(\xi) d\xi 
\end{align}
so that the arithmetic mean of the logarithm of the eigenvalues of $\pR$ in the Toeplitz form converges to the integral of $\log S$. It was shown by \cite{Adh13} that the eigenvalue spectrum $S(\xi)$ can be characterized in terms of only the geometric channel parameters by 
\begin{align} \label{eq:ID-30}
    S(\xi) = \frac{1}{2\Delta} \sum_{k\in [D\sin(-\Delta+\theta)+\xi,\; D\sin(\Delta+\theta)+\xi]} \frac{1}{\sqrt{D^2-(k-\xi)^2}}. 
\end{align}
As a consequence, we can accurately predict $\frac{1}{r}\log|\Lambdam_g|$ thanks to (\ref{eq:ID-29}) and (\ref{eq:ID-30}) in the one-ring model, avoiding the need for the calculation in (\ref{eq:SM-4}) and the eigendecompsition of the large-dimensional matrix $\pR_g$.


\subsection{Comparison with Independent Fading Channels}
\label{sec:IT-D}


\subsubsection{$r\ge K'$ case}

In this case, it turns out from comparing (\ref{eq:ID-6}) and (\ref{eq:ID-6b}) that transmit correlation diversity may increase the rate of growth of the asymptotic capacity of correlated fading BCs up to by $\log \mu=\log\frac{M}{K}$. This is because the eigen-beamforming gain can be as large as $\log \mu G$ by letting $K'$ strong eigenmodes take most of $\trace(\Lambdam_g)$ for all groups such that $\sum_{i=1}^{K'}\lambda_{g,i}\approx \trace(\Lambdam_g)=M$, as in (\ref{eq:ID-21}). This most favorable situation for the transmit correlation diversity gain may be too optimistic for all users to satisfy. Therefore, a less optimistic situation needs to be investigated in the sequel.

Let us first consider the case of $r=K'$ (i.e., $M=K$), in which the high-SNR behavior (\ref{eq:IT-14}) reduce to 
\begin{align}  \label{eq:IT-14b}
  \mathcal{C}^\text{sum}(P)   &=  M \log\frac{P}{M}+M\Bigg(-\gamma +\sum_{\ell=2}^{r}\frac{1}{\ell} \Bigg)\log e+\sum_{g=1}^G\log|\Lambdam_{g}|+o(1).
\end{align} 
By comparing this with $\mathcal{C}^\text{sum}_\text{iid}(P)$ in (\ref{eq:IT-50}),
transmit correlation diversity is shown to incur power loss in $\mathcal{C}^\text{sum}(P)$ due to the effective channel dimension reduction in (\ref{eq:SM-5}), represented by the offset between the second terms in (\ref{eq:IT-14b}) and (\ref{eq:IT-50}), i.e., $-M\big(\sum_{\ell=r+1}^{M}\frac{1}{\ell} \big)\log e$, and simultaneously it provides the eigen-beamforming gain of $\sum_{g=1}^G\log|\Lambdam_{g}|$. As a result, we observe a \emph{tradeoff} between the power loss and the power gain since $G$ is inversely proportional to $r$ for $M$ fixed.

In order to investigate the tradeoff and simplify the evaluation of the resulting gap between $\mathcal{C}^\text{sum}(P)$ and $\mathcal{C}^\text{sum}_\text{iid}(P)$, we upper-bound the asymptotic capacity behavior in (\ref{eq:IT-14b}) by letting 
\begin{align}  \label{eq:ID-25a}
  \lambda_{g,i}=\frac{M}{r}=G
\end{align} 
for all $(g,i)$. Then, the eigen-beamforming gain terms in (\ref{eq:IT-13}) and (\ref{eq:IT-14}) become 
\begin{align}  \label{eq:ID-25b}
  \sum_{g=1}^G\sum_{i=1}^{\min(r,K')}\log\lambda_{g,i} =\min(M,K)\log G
\end{align} 
which is in fact the upper bound in (\ref{eq:ID-20}) for $r<K'$.  Notice that, along with the  unitary structure, this condition (\ref{eq:ID-25a}) casts the corresponding correlated fading MIMO BC into parallel i.i.d. Rayleigh fading BCs with reduced channel dimension of $r$ each.
Using the approximation of the Harmonic number \cite{Con96}
\begin{align} 
   \sum_{\ell=1}^{n}\frac{1}{\ell} &=\gamma +\ln n +\sum_{m=2}^\infty \frac{\zeta(m,n+1)}{m} \nonumber \\
   &= \gamma +\ln n +\frac{1}{2n} -\frac{1}{12n^2} +\frac{1}{120n^4} +O({n^{-6}}) \nonumber
\end{align} 
where $\zeta(\cdot)$ is the Hurwitz zeta function,  we have 
\begin{align}  \label{eq:IT-20}
   \mathcal{C}^\text{sum}(P)-\mathcal{C}^\text{sum}_\text{iid}(P)& \le M\bigg(\frac{1}{2r} -\frac{1}{12r^2} -\frac{1}{2M} +\frac{1}{12M^2} \bigg)\log e +O(M^{-3})\nonumber \\
   & = \bigg(\frac{G-1}{2} -\frac{G^2-1}{12M}  \bigg)\log e +O(M^{-3})\nonumber \\
   &\approx \bigg(\frac{G-1}{2} \bigg)\log e \ \ \text{for large $M$ but fixed $G$.} 
\end{align}
This shows that, depending on the condition number of $\lambda_{g,i}$'s, the rate gap may be positive but marginal in the sense that it does not scale with $M$. By investigating the $r>K'$ case in a similar way with some manipulations, we can easily see that $\mathcal{C}^\text{sum}(P)-\mathcal{C}^\text{sum}_\text{iid}(P) \le K\big(\frac{G-1}{2} -\frac{G^2-1}{12M} -\sum_{\ell=r-K'+1}^r\frac{1}{\ell^2}\big)\log e+O(M^{-3})$, implying even the potential positive gain diminishes for $r> K'$.

Furthermore, it follows from replacing (\ref{eq:ID-21}) with (\ref{eq:ID-25b}) in the large system analysis that the above gain vanishes as $r $ and $K'$ increase with a fixed ratio $\mu\ge 1$, by noticing that the upper bound on the rate of growth in (\ref{eq:ID-6}) becomes now equivalent to (\ref{eq:ID-6b}). 
This shows that the performance loss due to the channel dimension reduction completely compensates the eigen-beamforming gain in this case. Consequently, our asymptotic bounds on the capacity behaviors are turned out to be useful to show that transmit correlation diversity in general provides no capacity gain for $r\ge K'$. Notice that we cannot get the above observations by directly using the previous results including (\ref{eq:IT-43}).

\subsubsection{$r< K'$ case}

We first consider the case where $r< K'$ but not $r \ll K'$.
In this case, we cannot accurately predict the impact of transmit correlation due to the lack of tightness of our asymptotic bounds in (\ref{eq:IT-13}) and (\ref{eq:ID-5}). However, if the partial cooperation is allowed, the high-SNR capacity of correlated fading BCs can approach close to that of the i.i.d Rayleigh fading point-to-point channel, depending on the condition number of $\lambda_{g,i}$'s again. 
In contrast, the independent fading case needs the full cooperation to achieve the same high-SNR capacity. But, this is much less feasible and the corresponding channel is not a BC any more. As a result, transmit correlation diversity is beneficial in this sense for $r< K'$ but not $r \ll K'$.


The more interesting case where $r \ll K'$ was already addressed by the authors in \cite{Nam13a}, but without sufficient exposition. 
It is well known from Sharif and Hassibi \cite{Sha05} that the sum capacity of the i.i.d. Rayleigh fading MIMO Gaussian BC scales like
$$  \Cc^\text{sum}_{\rm iid}(K) = M\log\frac{P}{M}+M\log\log K +o(1).$$
In the special case where all users have both the same SNR and the common transmit correlation matrix $\pR$ of full rank, the authors in \cite{Aln09} proved that the sum capacity scales like
$$  M\log\frac{P}{M}+M\log\log K +\log|\pR|+o(1)$$
where $\log|\pR|\le 0$ due to $\trace(\pR)=M$. The assumption that all users have the same transmit correlation is generally unrealistic in MU-MIMO downlink systems. 

Assuming the unitary structure where different groups of users have orthogonal eigenspaces, it was shown in \cite{Nam13a} that,  
for fixed $M$ and large $K$, the asymptotic sum capacity of correlated  Rayleigh fading MIMO BC is
\begin{align}  \label{eq:TC-3}
  \mathcal{C}^\text{sum}(K)=M\log\frac{P}{M}+M\log\log K +\sum_{g=1}^G\log|\Lambdam_{g}|+o(1)
\end{align}
where the detailed achievability proof is given in Appendix \ref{app:proof-3}. This shows that for large $K$ regime with correlated fading, there exists an additional term due to eigen-beamforming gain as well as the well-known multiuser diversity gain term $M\log\log K$. 
As an upper bound on the potential gain of transmit correlation diversity in the $r\ll K'$ regime, albeit the channel propagation for such gain is physically unrealistic, the following corollary was also presented: 
If the AS of group $g$, $\Delta_g$, is close to zero but Rayleigh fading is still valid,
then $ \limsup_{\Delta_g\rightarrow 0, \forall g}\Cc^\text{sum}(K)-\Cc^\text{sum}_{\rm iid}(K) = M\log M.$

In the former two cases of $r\ge K'$ and $r <K'$ but not $r\ll K'$, the eigen-beamforming gain was often completely compensated by the power loss due to the effective channel dimension reductio, yielding that transmit correlation does not help the capacity in general. In the last case, however, the power loss (corresponding to multiuser diversity gain reduction) \emph{vanishes} in the large $K$ regime as shown in Appendix \ref{app:proof-3}, while eigen-beamforming can still provide power gain of up to $M\log G$. This explains why the correlated fading case can significantly outperform the independent fading case in this regime.

Finally, we consider the partial cooperation for large $K'$ but not necessarily $r\ll K'$ and compare its performance with (\ref{eq:TC-3}).
\begin{corol} \label{cor-3}
Assuming the partial cooperation between the receivers within each group, we have
\begin{align}  \label{eq:IT-41}
  \Cc^\text{sum}(P,K) = M\log\frac{P}{M}+M\log K' +\sum_{g=1}^G\log|\Lambdam_{g}|+o(1).
\end{align}
\end{corol}

This can be easily verified with the high-SNR upper bound in (\ref{eq:IT-37a}) and the fact that 
\begin{align} 
   \mathbb{E} \big [\log\left|  \Wm_g \Wm_g^\ct \right|\big ]  \simeq r \log K', \ \text{ for large } K' \nonumber 
\end{align}
which follows from (\ref{eq:IT-36}) and (\ref{eq:ID-10}).
The partial cooperation is shown to provide the additional power gain of $M\big(\log K'-\log \log K\big)$ at high SNR for large $K$, compared to (\ref{eq:TC-3}) with no cooperation.

\subsection{Numerical Results and Summary}
\label{sec:IT-E}

In order to see if the foregoing results on the impact of transmit correlation deriven by imposing the unitary structure is still valid for realistic channels not assuming the ideal structure, we evaluate the ergodic sum capacity of correlated Rayleigh fading MIMO BCs generated by the one-ring model \cite{Shi00} in Section \ref{sec:SD}.

\begin{figure}
\vspace{-3mm}
\center  \includegraphics[scale=.8]{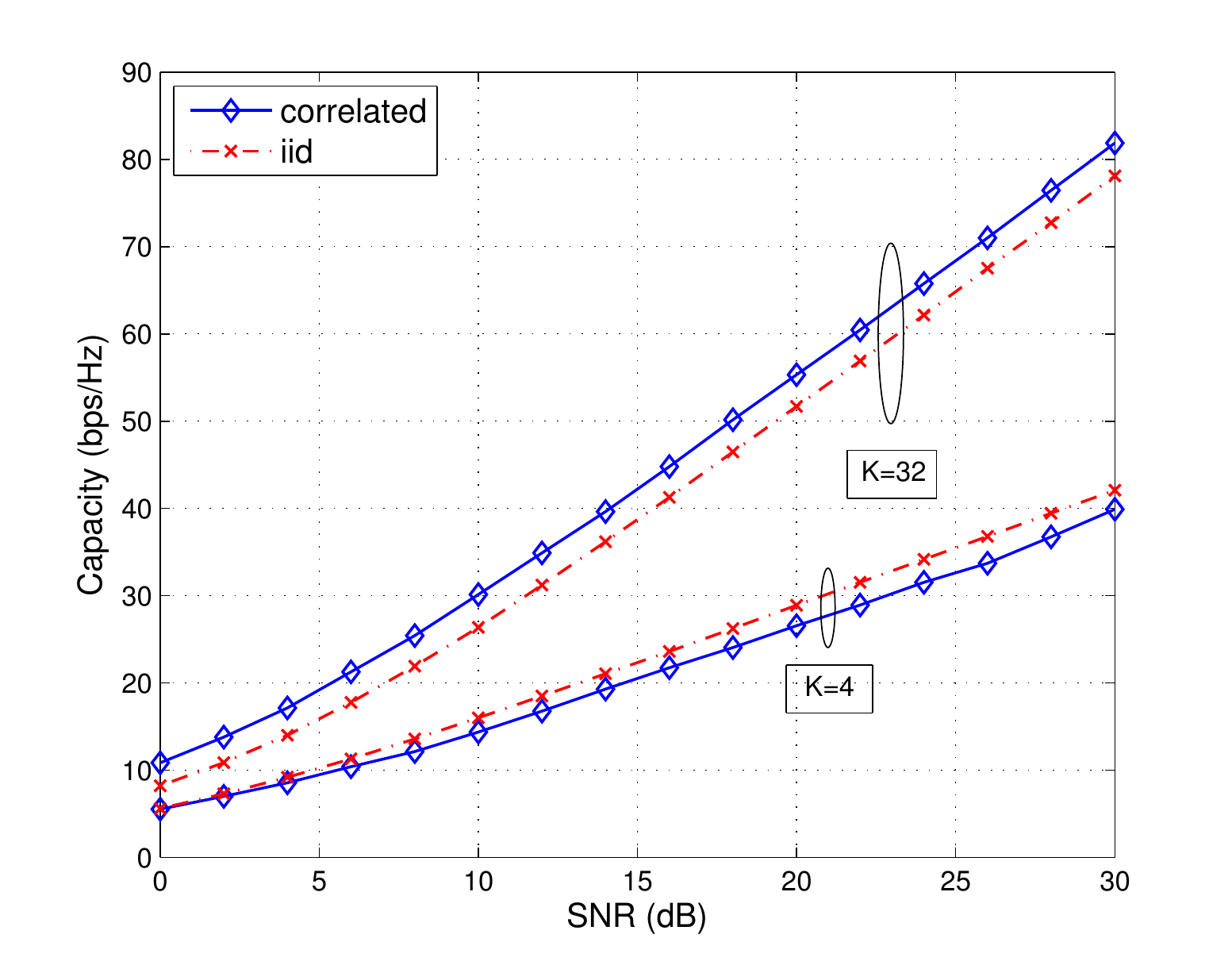}
  \caption{Sum capacity vs. SNR for $M=8$ and $\Delta_i\in[5^o,10^o]$. }\label{fig-4}
\vspace{-2mm}
\end{figure}

We consider the ULA with $D=1/2$ (half wavelength) and AoDs $\theta_i$ of users are uniformly distributed over the range $[-60^o, 60^o]$ with $\Delta_i$ uniformly distributed in the range $[5^o,20^o]$, where $\theta_i$ and $\Delta_i$ are AoD and AS of user $i$, respectively. The transmit correlation matrices $\pR_i$ of users are generated by the one-ring channel model (\ref{eq:SM-4}) and by the given distributions of geometric channel parameters.

Fig. \ref{fig-4} compares the i.i.d. and the correlated Rayleigh fading MIMO BCs in terms of the sum capacity versus SNR for different $K$, where $M=8$ and $\Delta_i\in[5^o,10^o]$. When $K=4$ ($r>K'$), the i.i.d. fading case has a larger capacity than the correlated fading case, as expected. When $K=32$ ($r<K'$), however, the capacity of the latter is larger than that of the former even if we did not assume the ideal unitary structure. This is because, in this regime, a ``semi-unitary" structure can be implicitly formed by  multiuser scheduling. This interesting result is also observed in the following evaluation. 

Fig. \ref{fig-7} shows the sum capacity versus the number of users of the i.i.d. and the correlated Rayleigh fading MIMO BCs for different $M$ and $\Delta_i$ when $r< K'$. Also, $M=4,8$ and $\Delta_i\in[2^o,5^o]$ or $[5^o,20^o]$. The rate gap between correlated and independent fading cases gets larger as the level of transmit correlation increases (i.e., $\Delta_i$ becomes smaller). In particular, when $K=10,000$ the rate gap is about $4.5$ bps/Hz for $M=4$ where the maximum possible gain of $M\log M$ is $8$, while it is around $12.5$ bps/Hz for $M=8$ where $M\log M=24$. Surprisingly, a large portion of the maximum potential gain of transmit correlation diversity is shown to be achievable for the realistic setup where no ideal structure of transmit correlations of users was assumed. In addition, it seems that the real asymptotic capacity scaling can be approximately predicted by the analysis in (\ref{eq:TC-3}), where we let $G=4$ for $M=8$ and $G=2$ for $M=4$ according to the two different distributions of AS $\Delta_i$ of users.

\begin{figure}
\vspace{-3mm}
\center  \includegraphics[scale=.8]{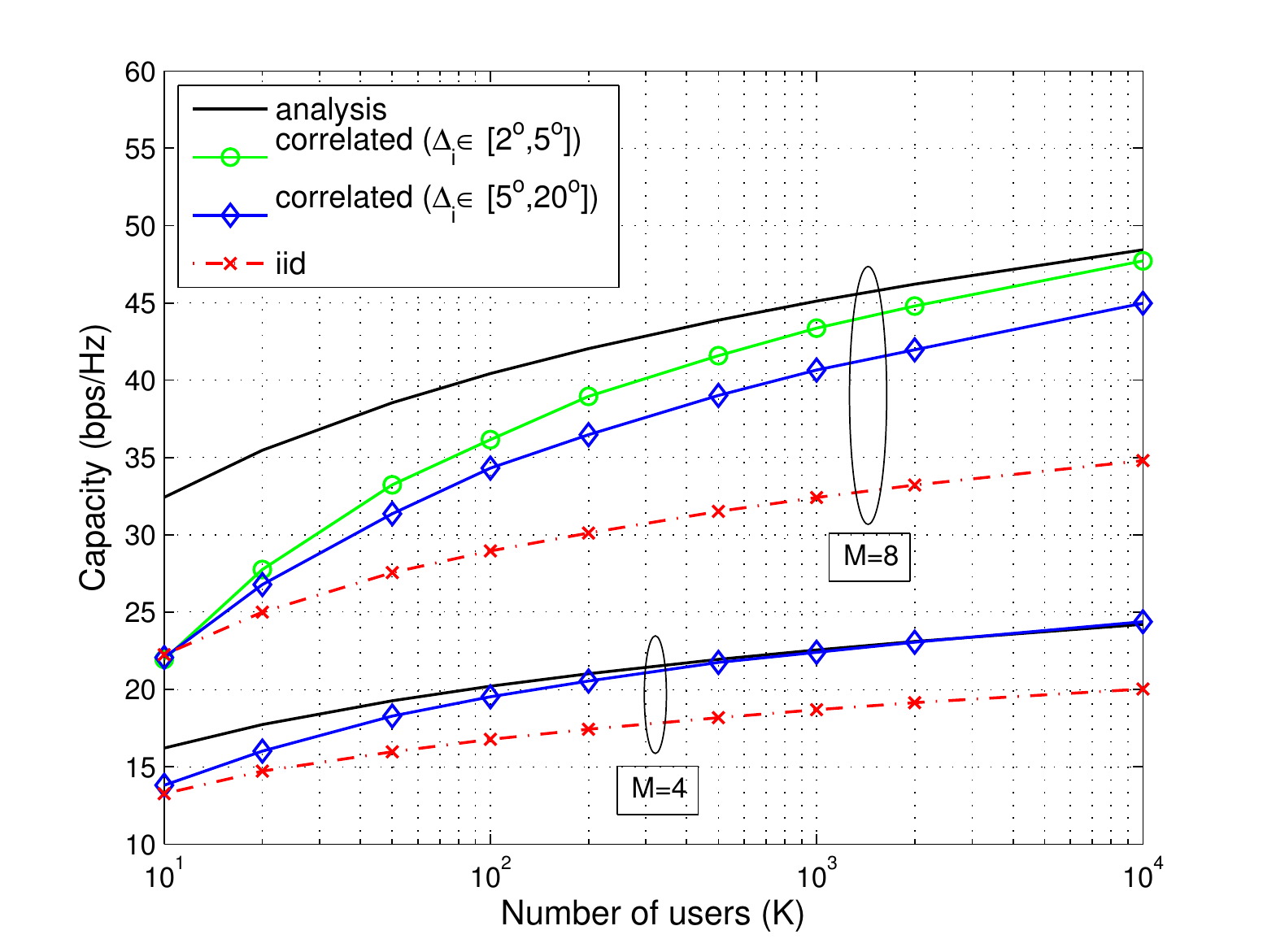}
  \caption{Sum capacity vs. the number of users for SNR $= 10$ dB and different ranges of angular spread.}\label{fig-7}
\vspace{-2mm}
\end{figure}

Assuming the unitary structure, the \emph{full-CSIT} capacity results in this section can be summarized as follows.
\begin{itemize}
\item For $r\ge K'$ (i.e, $M\ge K$), transmit correlation diversity may provide the potential power gain of up to $M\log\mu=M\log\frac{M}{K}$, compared to the i.i.d. Rayleigh fading MIMO BC. Given the less optimistic and more general condition in (\ref{eq:ID-25a}), however, correlated fading BCs can have {at most} marginal power gain at high SNR over the independent fading BC. In particular, for $r\gg K'$ or for large $r$ with $\mu>1$ fixed, even the marginal gain may vanishe. Therefore, transmit correlation has in general no beneficial impact on the capacity of MIMO BCs in this regime.
\item For $r<K'$ but not $r\ll K'$, we do not know exactly the impact of transmit correlation due to the lack of tightness of our asymptotic bounds. However, numerical results indicate that transmit correlation may have  capacity gain even when $r$ is not so smaller than $K'$ and the unitary structure is not available. 
In addition, the partial cooperation is sufficient to achieve the full-CSIT capacity of the point-to-point case. 
\item For $r \ll K'$, correlated Rayleigh fading BCs may have {much larger} sum-rate scaling than the i.i.d Rayleigh fading BC. We observe through numerical results that the gap becomes non-negligible for relatively small $\mu$, even if the unitary structure is not assumed again. 
\end{itemize}

Therefore, it turns out that transmit correlation diversity might be beneficial to the power gain of MIMO BCs for some special cases of $r\ge K'$ as well as for $r< K'$.

So far, we have assumed prefect CSIT with \emph{no cost}, for which {in general} we could not do better with transmit correlation diversity for $r\ge K'$. Notice that the typical scenario of large-scale MIMO belongs to this unfavorable case which includes $M\gg K$. Consequently, transmit correlation diversity could not improve the performance of large-scale MIMO systems at high SNR. It will be shown in the following section that this observation is not necessarily true for realistic pilot-aided systems, where CSIT is provided at the cost of downlink training.


\section{Fundamental Limits of Pilot-Aided MU-MIMO Systems}
\label{sec:FL}

In this section, we investigate asymptotic capacity bounds of pilot-aided MU-MIMO downlink systems, in which the resources for downlink training are taken into consideration. In general, realistic FDD systems make use of downlink common\footnote{In addition to the common pilot, some real-world systems like 3GPP long term evolution (LTE) employ downlink per-user (dedicated) pilot for coherent demodulation.} pilot and CSI feedback for downlink training \cite{Cai10}, while TDD systems employ uplink per-user pilot to exploit the uplink-downlink channel reciprocity and also need downlink per-user pilot for users to estimate their downlink channels for coherent demodulation. The pilot symbols for the latter can be shared by all scheduled users if $M$ is sufficiently large to invoke the law of large numbers (LLN). In this section, we do not consider the cost for CSI feedback and downlink per-user pilot, the latter of which will be discussed later in Section \ref{sec:PL}.

In FDD, a pilot-aided downlink system in the independent fading channel devotes the training phase of length $M_\text{iid}^*$ to allow users to estimate the $M_\text{iid}^*$-dimensional channel vectors, where $M_\text{iid}^* = \min\{M,K,\lfloor T_c/2\rfloor\}.$ 
Assuming that CSIT is acquired by delay-free and error-free feedback without channel estimation error, the high-SNR capacity of MU-MIMO downlink systems is upper-bounded by $M_\text{iid}^*(1-M_\text{iid}^*/T_c)\log \snr+O(1)$, as mentioned in (\ref{eq:intro-1}) and also \cite{Adh13}. Then, we have the following limit on the system multiplexing gain $$\lim_{\min\{M,K\}\rightarrow \infty} M_\text{iid}^*\left(1-\frac{M_\text{iid}^*}{T_c}\right)=\frac{T_c}{4}$$ for finite $T_c\in 2\mathbb{Z}^+$. This upper bound is also valid in TDD. For example, the number of scheduled users (i.e., $s=M_\text{iid}^*$ for the independent fading case) among the entire $K$ users is limited by uplink pilot overhead (also affected by $T_c$) in TDD large-scale MIMO systems \cite{Mar10}. 
If instantaneous feedback within coherence time $T_c$ is not possible, the impact of the resulting channel prediction error on the system multiplexing gain can be found in \cite{Cai10}.

As already pointed out, the above factor $T_c/2$ significantly limits the system performance for both $M$ and $K$ large. 
Noticing that this result holds true in the \emph{i.i.d. fading} channel, however, we will characterize some  fundamental performance limits in \emph{correlated fading} channels with the notion of transmit correlation diversity. In the sequel, suppose that the unitary structure is attained as before and that $T_c$ is finite.


\subsection{Training Overhead Reduction}
\label{PS-0}

\subsubsection{FDD (Pre-beamformed Pilot)}

The common pilot is in general isotropically transmitted, since it has to be seen by all users. We first consider a simple training scheme for FDD systems, where the downlink common pilot signal $\pX_g^\text{dl}$ for group $g$ is given by the pre-beamforming matrix $\pB_g$ as follows:
$$\pX_g^\text{dl}=\pB_g \rho_\text{tr}\pI_r$$
where $\rho_\text{tr}$ indicates the power gap between the training phase and the communication phase. Thanks to the unitary structure, we can let $\pB_g=\pU_g$ and the received pilot signal matrix for group $g$ is given by
\begin{align}  \label{eq:PS-1}
  \pY^\text{dl}_g = \pH_g^\ct \pX^\text{dl} +\pZ_g^\text{dl} = \rho_\text{tr}\textsf{\pH}_g^\ct +\pZ_g^\text{dl} 
\end{align}
where $\pX^\text{dl} =\sum_{g=1}^{G} \pX_g^\text{dl}$. This indicates that $G$ \emph{pre-beamformed pilot} signals $\px^\text{dl}_{g,i}$, $\forall g$, where $\px^\text{dl}_{g,i}$ is the $i$th column of $\pX^\text{dl}_{g}$, can be multiplexed and transmitted through a single pilot symbol and hence the overall common pilot signal $\pX^\text{dl}$ consumes only $r$ symbols, reduced by a factor of $G$.

Based on the above noisy observation of the pilot signal, each user in group $g$ can estimate the effective channel $\textsf{\ph}= \pU_g^\ct \ph$, which is unitarily equivalent to ${\ph}$ in (\ref{eq:SM-2}) under the unitary structure, as shown in Sec. \ref{sec:TCD}. Therefore, the proposed common pilot incurs no loss due to pre-beamforming, as if it were a conventional pilot signal isotropic to users in each group.
A generalization of the above scheme was already given in \cite{Adh13}, which evaluated the impact of the noisy CSIT on the JSDM performance by letting $\pX_g^\text{dl}=\pB_g\pU^\text{dl}$ with $\pU^\text{dl}$ being a scaled unitary matrix of size $r\times r$,  thereby making the downlink common pilot signal for each of antennas spread over $r$ pilot symbols. However, the work did not consider an optimization of the system degrees of freedom taking into account the cost for downlink training dimension.

\subsubsection{TDD}

 The same line of thought can be naturally applied to the TDD case with receive beamformer $\pU_g^\ct$ for the uplink per-user pilot. To be specific, 
the received pilot signal matrix for TDD systems can be given by
$$\pY^\text{ul} = \sum_{g=1}^G \pH_g  \rho_\text{tr}\pI_{K'} +\pZ^\text{ul}. $$
By receive beamforming, i.e., multiplying from the left by $\pU_g^\ct$ for group $g$, we have
\begin{align}  \label{eq:PS-2}
  \pY^\text{ul}_g = \pU_g^\ct\pY^\text{ul} = \rho_\text{tr}\textsf{\pH}_g +\tilde{\pZ}_g^\text{ul} 
\end{align}
where $\tilde{\pZ}_g^\text{ul} =\pU_g^\ct{\pZ}^\text{ul}$.  The uplink per-user pilot signal for all $K$ users consumes only $K'$ symbols, reduced by a factor of $G$ again. As a result, we can obtain the pilot saving not only in FDD systems but also in TDD systems, where the unitary structure is uplink-downlink reciprocal. Notice that such a pilot saving is also valid for MIMO MAC, i.e., MU-MIMO uplink systems. In \cite{Yin13}, a similar idea to the unitary structure was differently used to eliminate the pilot contamination effect in the multi-cell framework instead of reducing the overhead for uplink per-user pilot in each single cell. 

In terms of training, we compare JSDM with space-division multiple access (SDMA) referring to a wide class of beamforming schemes that depend on \emph{instantaneous} channel orthogonality (spatial separation) between users (e.g., \cite{Vis02,Cai03,Spe04,Sha05,Yoo06}). 
SDMA requires orthogonal common pilots since, unlike user data streams, the common pilot cannot be multiplexed by SDMA. 
In contrast, the virtual sectorization of JSDM allows a common pilot resource to be reused over all orthogonal groups (virtual sectors), thereby yielding a remarkable saving in the training phase. 
This can provide an insight for designing downlink training schemes for large-scale MIMO. For example, refer to a recent work in \cite{Lee14}. The training overhead reduction makes a crucial impact on the multiplexing gain of pilot-aided systems, as will be seen in the following subsection.

\subsection{Pilot-Aided System I}
\label{PS-1}

Following the pilot-aided scheme proposed in \cite{Zhe02}, suppose that we use $Q$ of $M$ BS antennas in the communication phase for FDD systems\footnote{In the TDD case, it suffices to suppose that we schedule $Q$ of $K$ users, where $Q\le K$, and to optimize the degrees of freedom with respect to $Q$ taking into account the uplink pilot overhead.}, where $Q\le M$.   Assuming the unitary structure, the total number of degrees of freedom for communication is upper-bounded by  
\begin{align} \label{eq-01}
   \min\{Q,K\}\left(T_c-\Big\lceil\frac{Q}{G}\Big\rceil\right) 
\end{align}
where $Q/G$ is due to the above training overhead reduction. Thus, we need to devote only $\lceil\frac{Q}{G}\rceil$ channel uses to estimate $Q$-dimensional channel vectors in the training phase. We call this \emph{pilot-aided system I} in this work. It is easy to show that the optimal number of transmit antennas to use is 
\begin{align} \label{eq-0}
   M^*=\min\left\{M,K,\Big\lfloor \frac{T_cG}{2}\Big\rfloor\right\}
\end{align}
where the subscript $\text{p1}$ indicates pilot-aided system I, yielding the pre-log factor 
$M^*\big(1-\frac{M^*}{T_cG}\big)$.
Therefore, we obtain the fundamental limit in (\ref{eq:intro-4}) and the limit on multiplexing gain
$$\lim_{\min\{M,K\}\rightarrow \infty} M^*\left(1-\frac{M^*}{T_cG}\right)=\frac{T_cG}{4}$$
 for finite $T_cG\in 2\mathbb{Z}^+$.
 
It can be seen from (\ref{eq-0}) that, for both $M$ and $K$ large, exploiting $G$ degrees of transmit correlation diversity  can increase the system multiplexing gain up to by a factor of $G$. 
It is evident that as long as the degrees of transmit correlation diversity is sufficiently large such that $G\ge {2\min\{M,K\}}/{T_c}$, the optimal number of transmit antennas $M^*$ is not affected any longer by the coherence time interval $T_c$. As a consequence, the system multiplexing gain is not saturated and keeps growing as $\min\{M,K\}$ increases. 
The following example compares the upper bound (\ref{eq:intro-1}) on the system multiplexing gain for the independent fading case and the new upper bound (\ref{eq:intro-4}) for the correlated fading case, when $T_c$ is taken from real-life cellular systems.

\begin{ex}
Let $T_c$ take either 32 long-term evolution (LTE)  symbol duration \cite{LTE12} (approximately 60 km/h) or 100 symbol time (19 km/h).  
Also, suppose that the unitary condition is attained such that $G=4$ and $G=8$. Fig. \ref{fig-1} shows the Zheng-Tse upper bound, $M_\text{iid}^*(1-M_\text{iid}^*/T_c)$, and the new bound $M^*(1-M^*/T_cG)$ on the system multiplexing gain as $\min\{M,K\}$ increases. It can be seen that exploiting transmit correlation diversity can increase the multiplexing gain up to by a factor of $4$ and $8$ for $G=4$ and $G=8$, respectively.
\end{ex}

So far, we have fixed the number of degrees of transmit correlation diversity, $G$. We now turn our attention to the case where as $M\rightarrow \infty$, $G$ also grows such that the ratio $G/M$ is not vanishing (i.e., {bounded}). In practical systems, we generally consider a  large-scale array in the high carrier frequency $f_c$ due to the space limitation of large-scale array and the fact that the wavelength is inversely proportional to $f_c$. So, it is fairly reasonable to increase $M$ proportionally as $f_c$ grows and hence to let $M$ depend on $f_c$. It was observed, e.g., in mm-Wave channels \cite{Zha10}, that the higher  $f_c$, the smaller number of strong multipaths the receivers experience due to higher directionality. This sparsity of dominant multipath components is also verified by mm-Wave propagation measurement campaigns \cite{Rap13}. Thus, the high transmit correlation diversity is attainable in the high $f_c$ case. Then, as both $M$ and $f_c$ grow, $r$ may remain unchanged such that $G/M$ is fixed.

In what follows, using the results in the previous section, we refine the $O(1)$ term in (\ref{eq:intro-4}) first in the large $G$ regime and then in the large $r$ regime. 
Let $$r_\text{p1}=\frac{M^*}{G}\; \text{ and } \; \mu_\text{p1}=\frac{M^*}{K}$$ be fixed, where $G$ is assumed to divide $M^*$ such that $r_\text{p1} \in \mathbb{Z^+}$ for simplicity. Given a finite coherence time interval $T_c$, also let  
$$\nu=\frac{M^*}{T_cG}=\frac{r_\text{p1}}{T_c}.$$  
In this scenario, $G$ is taken to infinity along with $M$ but both $r$ and $K'$ are finite, unlike Theorem \ref{thm-8}. Therefore, we shall make use of (\ref{eq:ID-2}) instead of (\ref{eq:ID-4}) in Lemma \ref{lem-1} and then we can apply Theorem \ref{thm-2} in the sequel.

Denote by $\mathcal{C}^\text{sum}_\text{p1}(P,M^*,\upsilon)$ the high-SNR capacity of pilot-aided system I for $M^*$ and $\upsilon$ large, where $\upsilon = r \text{ or } G$. 
Assuming that the perfect CSIT is provided by an ideal (i.e., delay-free and error-free feedback) uplink with no channel estimation error in FDD and neither calibration error nor pilot contamination in TDD, respectively, $\mathcal{C}^\text{sum}_\text{p1}(P,M^*,\upsilon)$ is simply given by
\begin{align} 
  \mathcal{C}^\text{sum}_\text{p1}(P,M^*,\upsilon) &= (1-\nu)\; \Cc^\text{sum}(P,M^*,\upsilon). \nonumber 
\end{align}

\begin{thm} \label{thm-4}
Suppose the perfect CSIT on ${\textsf{\pH}}_g$ and the unitary structure. As $M\rightarrow \infty$, for $\mu< 1$,  the high-SNR capacity of the corresponding correlated fading MIMO BCs scales linearly in $M^*$ with the ratio
\begin{align} \label{eq:ID-15}
  \frac{\mathcal{C}^\text{sum}_\text{p1}(P,M^*,G)}{M^*} &= (1-\nu)\left\{{\log\frac{P}{r_\text{p1}}}+\log e\Bigg(-\gamma+\sum_{\ell=2}^{K'}\frac{1}{\ell}+\Big(\frac{1-\mu_\text{p1}}{\mu_\text{p1}}\Big)\sum_{\ell=(1-\mu_\text{p1})K'+1}^{K'}\frac{1}{\ell}\Bigg)+c_{G,1}\right\} +o(1).
\end{align}
where the constant $c_{G,1}$ is bounded by $$\log\frac{\mu}{\zeta} \le c_{G,1}\le 0.$$

For $\mu\ge 1$, the high-SNR capacity scales linearly in $M^*$ with the ratio
\begin{align}  \label{eq:ID-13b}
  \frac{\mathcal{C}^\text{sum}_\text{p1}(P,M^*,G)}{M^*} &= (1-\nu)\left\{{\log\frac{P}{r_\text{p1}}}+\log e\Bigg(-\gamma+\sum_{\ell=2}^{K'}\frac{1}{\ell}\Bigg)+c_{G,2}\right\} +o(1)
\end{align}
where $c_{G,2}$ is bounded by $$-\log\zeta \le c_{G,2}\le \log\mu.$$
\end{thm}

\begin{IEEEproof} 
For $\mu < 1$ and $\nu\le 1/2$, we have $M^*=M, r_\text{p1}=M/G=r, \mu_\text{p1}=\mu$. In this case, we use (\ref{eq:IT-13}) because it becomes loose only when $\mu$ is small. Then, the ratio at which the high-SNR capacity increases in the large $G$ regime as $M\rightarrow \infty$ is
\begin{align}  \label{eq:ID-13}
   \frac{\mathcal{C}^\text{sum}(P,M,G)}{M} &=  \log\frac{P}{M}+\log e\sum_{g=1}^G \frac{r}{M} \Bigg(-\gamma +\sum_{\ell=2}^{K'}\frac{1}{\ell}+\frac{K'-r}{r}\sum_{\ell=K'-r+1}^{K'}\frac{1}{\ell} \Bigg)\nonumber \\
   &\ \ \ \ \ +\frac{1}{M}\sum_{g=1}^G\log|\Lambdam_g|+\frac{c_{P,1}}{M}+o(1) \nonumber \\
   &= {\log\frac{P}{r}}+\log e\Bigg(-\gamma+\sum_{\ell=2}^{K'}\frac{1}{\ell}+\Big(\frac{1-\mu}{\mu}\Big)\sum_{\ell=(1-\mu)K'+1}^{K'}\frac{1}{\ell} \Bigg)+c_{G,1}+o(1) 
\end{align}
where we used (\ref{eq:ID-20}) and the uniform boundedness of the eigenvalues $\lambda_{g,i}$ in (\ref{eq:SM-1b}). 
When $\nu > 1/2$, the rate of growth for the $\mu < 1$ case can be similarly obtained by noticing $M^*=\frac{T_cG}{2}$ and $r_\text{p1}=\frac{T_c}{2}$.
Therefore, for these two cases in the large $G$ regime with $r_\text{p1}$ fixed, we get (\ref{eq:ID-15}).

For $\mu \ge 1$ and $\nu\le 1/2$, noticing that $M^*=K, r_\text{p1}=K/G=K', \mu_\text{p1}=1$ and using (\ref{eq:IT-14}), (\ref{eq:ID-21}), and (\ref{eq:SM-1b}), 
we get 
\begin{align}  \label{eq:ID-14}
  \frac{\mathcal{C}^\text{sum}(P,K,G)}{K}   &=  \log\frac{P}{K}+\log e\sum_{g=1}^G \frac{K'}{K}\Bigg(-\gamma +\sum_{\ell=2}^{K'}\frac{1}{\ell}\Bigg)+\frac{1}{K}\sum_{g=1}^G \sum_{i=1}^{K'}\log \lambda_{g,i}+\frac{c_{P,2}}{K}+o(1) \nonumber \\
  &=  {\log\frac{P}{K'}}+\log e\Bigg(-\gamma+\sum_{\ell=2}^{K'}\frac{1}{\ell}\Bigg) +c_{G,2}+o(1)
\end{align}  
where we used (\ref{eq:ID-21}) and (\ref{eq:SM-1b}) as before.
When $\nu > 1/2$, the ratio for the $\mu \ge 1$ case can be  obtained again by noticing $M^*=\frac{T_cG}{2}$ and $r_\text{p1}=\frac{T_c}{2}$. Then, we obtain (\ref{eq:ID-13b}) for $\mu \ge 1$.
\end{IEEEproof}


The following result shows the \emph{most optimistic} gain of transmit correlation diversity in the limit of $\Delta_g\rightarrow 0$ for all $g$, which we provide as a capacity upper bound even though this channel assumption seems unrealistic. 

\begin{corol} \label{cor-4}
For $\mu=1$ and $T_c \ge 2$, as $\Delta_g\rightarrow 0$ and $M\rightarrow \infty$ with $\mu$ and $\nu$ fixed, the high-SNR capacity of the pilot-aided system I scales linearly in $M$ with the maximum possible ratio  
\begin{align} \label{eq:ID-1}
  \lim_{M \rightarrow \infty}\limsup_{\Delta_g\rightarrow 0}\frac{\mathcal{C}^\text{sum}_\text{p1}(P,M,G)}{M} = \left(1-T_c^{-1}\right) \log\frac{P}{e}. 
\end{align}
\end{corol}

To prove this, we first notice that the condition of (\ref{eq:intro-4}) can be restated as $T_c \ge 2 \min(r,K')$ in this case and hence the sufficient condition is guaranteed just for $T_c \ge 2$, since $r\rightarrow 1$ as $\Delta_g\rightarrow 0$. Using this and (\ref{eq:ID-15}), (\ref{eq:ID-1}) immediately follows with $\nu=T_c^{-1}$.
It is remarkable that if the unitary structure is attained with $T_c$ sufficiently large and $\Delta_g$ sufficiently small, the high-SNR capacity of MU-MIMO systems approaches to the full-CSI capacity in (\ref{eq:ID-23}).  The systems of interest are \emph{scalable} in $\min(M,K)$ and also the user throughput does not vanish any longer unless $K\gg M$, unlike (\ref{eq:intro-1}). We point out that Corollary \ref{cor-4} could be obtained not just by pilot saving but also by the power gain due to eigen-beamforming.

In the large $r$ regime where $r$ goes to infinity while $G$ fixed, we obtain the following result by using Theorem \ref{thm-8}.
For $\mu_\text{p1}\le 1$ 
\begin{align}  \label{eq:ID-16}
  \frac{\mathcal{C}^\text{sum}_\text{p1}(P,M^*,r)}{M^*} =  (1-\nu)\left\{ \log \frac{P}{e\mu_\text{p1}}+\Big(\frac{1-\mu_\text{p1}}{\mu_\text{p1}}\Big)\log\frac{1}{1-\mu_\text{p1}}+c_{\text{p1},1}\right\} +o(1)
\end{align}
where $\log \frac{\mu_\text{p1}\lambda_\text{min}}{G} \le c_{\text{p1},1}\le 0$.
Note that according to (\ref{eq-0}), the case of $\mu_\text{p1}>1$ does not happen in pilot-aided system I, since we make use of only $K$ transmit antennas regardless of how large $M$ is, i.e., $M^*=K$. This phenomenon in system I may cause a nontrivial rate loss for $M> K$, as will be discussed in the next subsection.

\begin{figure} 
\center \includegraphics[scale=.8]{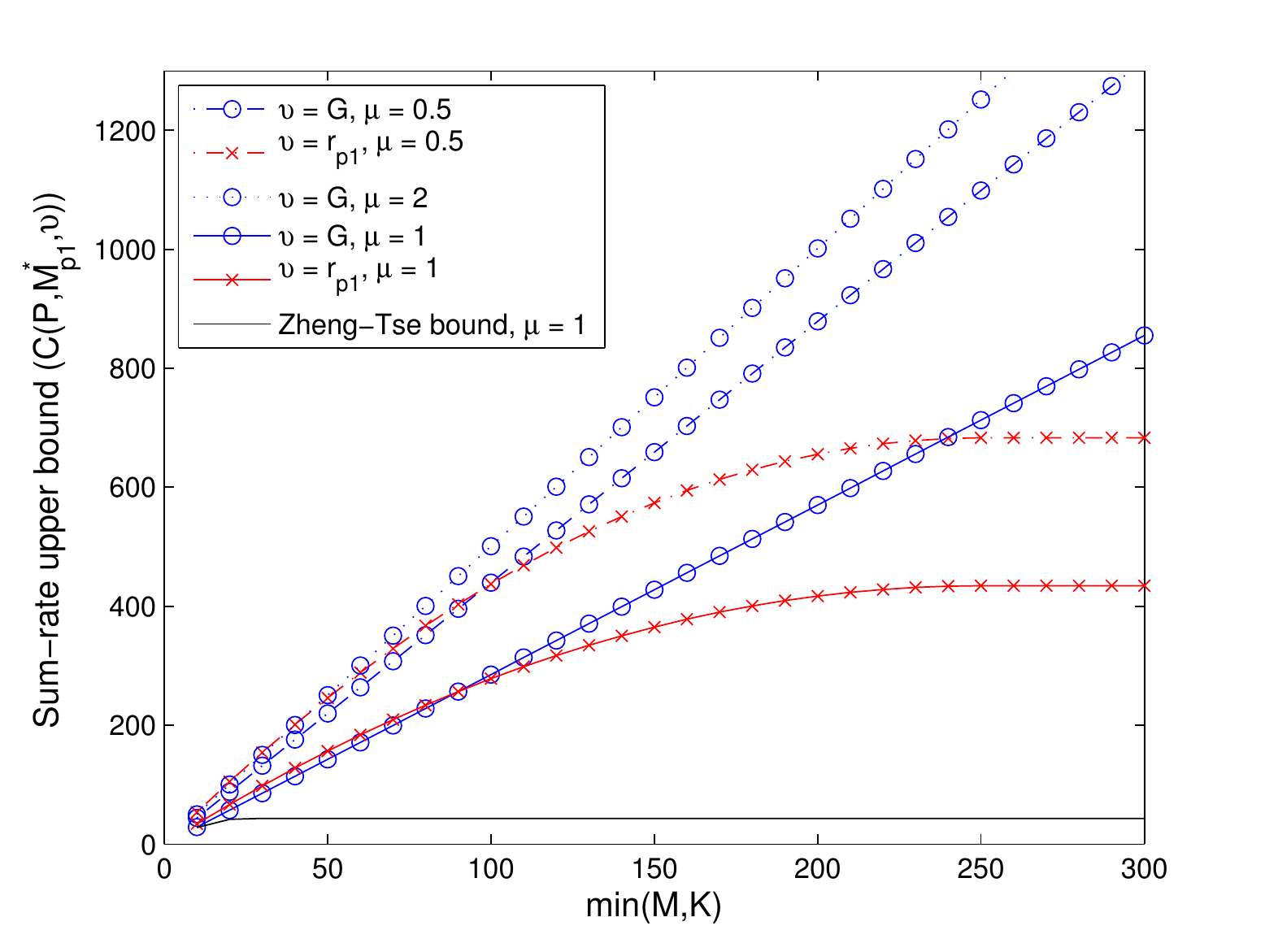} 
\caption{Asymptotic sum-rate upper bound curves versus $\min(M,K)$ in pilot-aided system I at $P=30$ with $T_c=50$, where $r_\text{p1}=10$ when $G$ is large ($\upsilon=G$), and  $G=10$ when $r_\text{p1}$ is large ($\upsilon=r_\text{p1}$).} \label{fig-2}
\end{figure}

Fig. \ref{fig-2} shows sum-rate upper bounds in (\ref{eq:ID-15}), (\ref{eq:ID-13b}), and (\ref{eq:ID-16}) on the asymptotic capacity for different system parameters in pilot-aided systems I. The rate of growth of the Zheng-Tse bound (\ref{eq:intro-1}) for large $M$ is given by $M_\text{iid}^*(1-M_\text{iid}^*/T_c)\log \frac{P}{e}+o(1)$.
For large $G$ and fixed $r_\text{p1}$, the system multiplexing gain grows linearly with $\min\{M,K\}$, whereas this is not the case with large $r_\text{p1}$ and fixed $G$. 
To understand the large rate gap between $\mu=1$ and $\mu=2$, recall the optimistic eigen-beamforming gain of up to $\log \mu G$ in Section \ref{sec:IT-D} and that a dual MAC is equivalent at high SNR to the corresponding MIMO point-to-point channel with $K$ transmit antennas and $M$ receive antenna. The equivalent MIMO channel is well understood to have a logarithmic power gain scaling with $M$ due to receive beamforming. For $\mu=0.5$ case, the large rate gap from $\mu=1$ is because the upper bound was given by allowing the partial cooperation within each group. 
Finally, for large $r_\text{p1}$ but fixed $G$, the two cases of $\mu=1$ and $\mu=2$ collapse into the red solid line. This is due to the fact that pilot-aided system I considers only multiplexing gain but not power gain, which will be addressed in the following subsection.

\subsection{Pilot-Aided System II}

In the large $M$ regime, the $\mu > 1$ case may be more frequently encountered in realistic systems, which is also the typical scenario of large-scale MIMO.  
We introduce a new pilot-aided system to address the foregoing issue for this case with $r$ large but $G$ fixed. 
In contrast to pilot-aided system I, in which only $K$ transmit antennas are used by letting $M^*=K$ when $M\ge K$, we shall allow in the new system referred to as \emph{pilot-aided system II} to use more than $K$ transmit antennas, even though the degrees of freedom is certainly at most $K$. By doing so, we may obtain a noticeable power gain suggested by (\ref{eq:ID-6}) due to transmit correlation diversity which compensates the increase in channel uses required for downlink training. To understand this, notice that using more than $K$ antennas has less impact on the system multiplexing gain as $G$ and/or $T_c$ grows, as shown in (\ref{eq:intro-4}).

\begin{figure} 
\center\includegraphics[scale=.8]{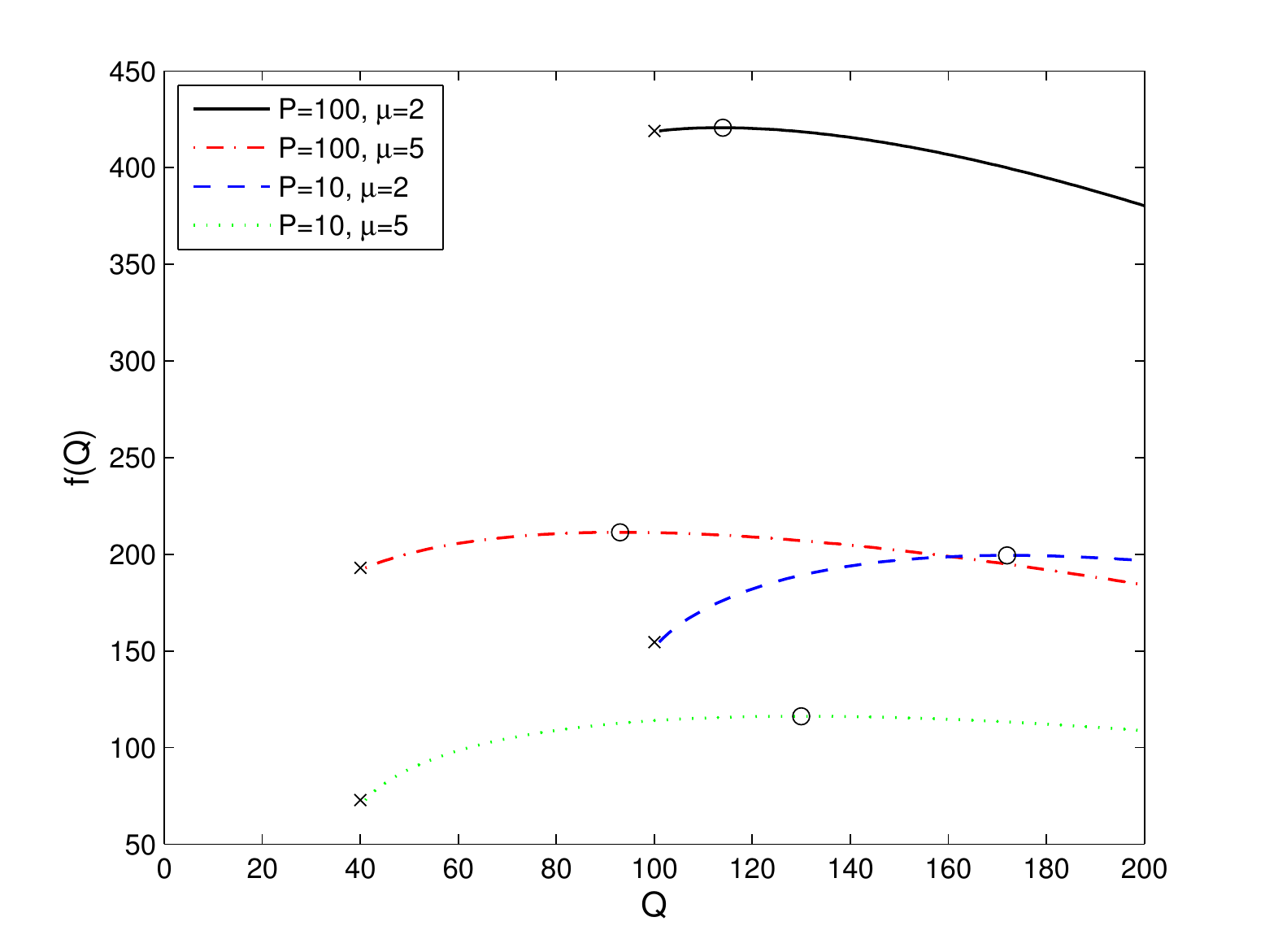} 
\caption{Values of $f(Q)$ versus the number of transmit antennas to use, $Q$, in pilot-aided system II when $M> K$ (i.e., $\mu > 1$), where $M=200$, $T_c=64$, and the `o' indicates the optimum numbers of transmit antennas, $M_\text{p2}^*$, and the `x' indicates $M^*$.} \label{fig-3}
\end{figure} 

To take into account the additional power gain from using more than $M^*$ transmit antennas in pilot-aided system II, we replace the optimization problem in (\ref{eq-01}) with the following one based on the upper bound in (\ref{eq:ID-6}).  
\begin{align} \label{ID-18}
   M_\text{p2}^* =\argmax_{Q} f(Q)
\end{align}
subject to $M^*\le Q\le M$, where $f(Q) = M^*\left(T_c-\lceil\frac{Q}{G}\rceil\right)\log \frac{P}{e}\frac{Q}{K}+(\frac{Q}{K}-1)\log\frac{Q}{Q-K} +\log\frac{Q}{K}$ with $M^*$ (the maximum number of degrees of freedom for the communication phase) unchanged. The high-SNR capacity of pilot-aided system II for $\mu>1$ scales linearly in $K$ with the ratio
\begin{align}  \label{eq:ID-17}
  \frac{\mathcal{C}^\text{sum}_\text{p2}(P,M_\text{p2}^*,r)}{K}  &= (1-\nu_\text{p2})\left\{ \log \frac{\mu_\text{p2}P}{e}+(\mu_\text{p2}-1)\log\frac{\mu_\text{p2}}{\mu_\text{p2}-1}+c_{\text{p2},2}\right\} +o(1)
\end{align}
where $\nu_\text{p2}=\frac{M_\text{p2}^*}{T_cG}$, $\mu_\text{p2}=\frac{M_\text{p2}^*}{K}$, and $\log \frac{\lambda_\text{min}}{G} \le c_{\text{p2},2}\le \log \mu_\text{p2}$.

Fig. \ref{fig-3} shows the optimum number of transmit antennas, $M_\text{p2}^*$, for different $P$ and $\mu$ with $M=200$. Here, $M^*=K=40$ for $\mu=5$ and $M^*=K=100$ for $\mu=2$. Therefore, if we consider not only the system multiplexing gain but also the power gain due to eigen-beamforming, the optimum values of $M_\text{p2}^*$ are shown to be quite different from $M^*$. We can also see that the resulting rate gap is reduced as $P$ increases for $T_c=64$.
Fig. \ref{fig-5} compares the asymptotic sum-rate upper bounds of pilot-aided system I and II when $T_c=32$ and $T_c=128$. It is shown that the rate gap gets larger as $T_c$ increases, since, for large $T_c$, the extra overhead due to training more than $K$ antennas reduces, as mentioned earlier.

\begin{figure} 
\center\includegraphics[scale=.8]{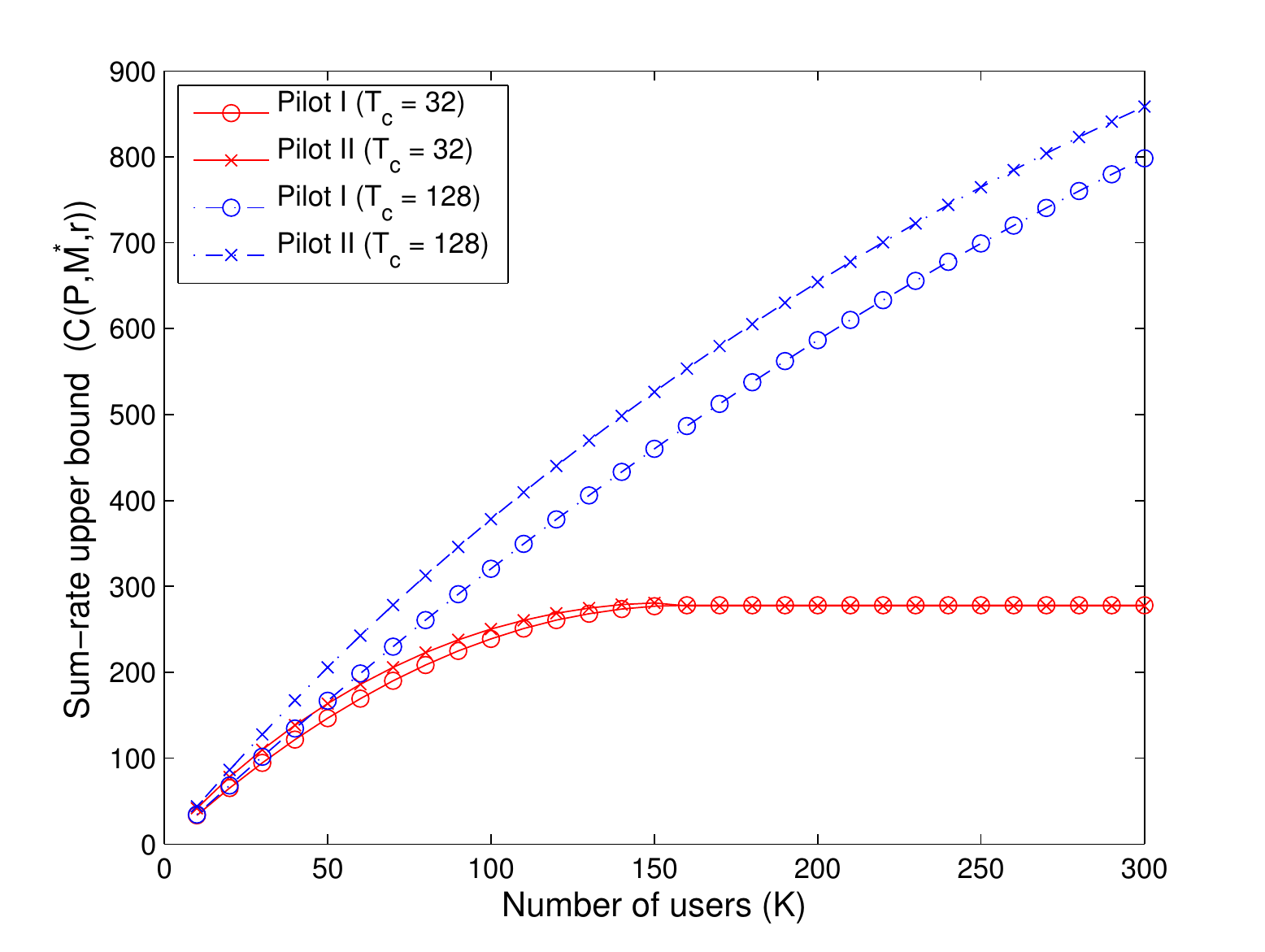} 
\caption{Asymptotic sum-rate upper bounds of two pilot-aided systems where $\mu=2$, $G=10$, and $P= 30$.} \label{fig-5}
\end{figure} 

\begin{remark}
So far, we have assumed $T=1$ such that channel covariances of all users associated to the BS satisfy a single unitary structure, which is in general unrealistic. If we extend to the case of multiple classes as shown in Fig. \ref{fig-1}, then the overhead of the pilot design in Sec. \ref{PS-0} is subject to increase by a factor of $T$, e.g., replacing the pre-log factor in (\ref{eq:intro-4}) with 
\begin{align} \label{eq-02}
   M^*\left(1-\frac{M^*T}{T_cG}\right)
\end{align}
where $M^*=\min\left\{M,K,\lfloor \frac{T_cG}{2T}\rfloor\right\}$.
This may undermine the potential gain of transmit correlation diversity, yielding the system design guideline that there exists a tradeoff between the number ($T$) of classes and the system performance so that $T$ should be less than the number ($G$) of degrees of transmit correlation diversity and it needs to be restricted as small as possible. 
\end{remark}

\section{Performance Limit of Large-Scale MIMO for TDD systems}
\label{sec:PL}

In this section, we consider conventional TDD large-scale MIMO systems based on the instantaneous channel reciprocity but not exploiting transmit correlation diversity to understand their performance limits guided by the results in this work. One of the essential building blocks for realistic large-scale MIMO systems is the feasibility in terms of training overhead. Therefore, we need to more carefully look at the following two aspects of the training cost that may significantly affect the system performance.

First, we have considered only uplink per-user (dedicated) pilot in TDD systems, but downlink per-user pilot should be taken into account as well. This is because even if we employ the maximal-ratio single-user beamforming scheme in \cite{Mar10} and the transmitter is assumed to perfectly know the channel vectors of users, each receiver requires a reference signal of the received quadrature amplitude modulation (QAM) constellation for coherent detection. Provided that the number of antennas is large enough to invoke LLN enabling the \emph{favorable propagation condition} with perfect CSIT (i.e., neither pilot contamination effect nor large-scale array calibration errors), it is sufficient to use only a common resource shared by all downlink per-user pilots. Otherwise, we should make use of orthogonal resources to prevent a severe performance degradation due to interference between non-orthogonal downlink per-user pilots. This performance degradation would remain significant until $M$ reaches a threshold denoted by $N_\text{LLN}$. 
Secondly, we have assumed a frequency flat (narrow-band) channel. In \cite{Mar10}, a \emph{frequency smoothness interval} was used to take into account the frequency selectivity in wide-band channels. To the best of our knowledge, the impact of the downlink per-user pilot overhead to the system multiplexing gain has not been addressed in the large-scale MIMO context.

Many of current commercial cellular systems employ the following pilot design:  
Downlink training for FDD systems consists of downlink common and per-user pilots, while the TDD downlink training requires uplink per-user pilot and downlink per-user pilot\footnote{In the LTE context \cite{LTE12}, the downlink common, downlink per-user, and the uplink per-user pilots correspond to CSI reference signal (CSI-RS), demodulation reference signal (DM-RS), and sounding reference signal (SRS), respectively.}. Downlink common and uplink per-user pilots are used for the BS to acquire CSIT in FDD and TDD, respectively. For both duplex modes, downlink per-user pilot is required for coherent demodulation. The downlink per-user pilot requires in general much more pilot symbols\footnote{For the LTE-advanced system with $M=4$, CSI-RS consumes at most $4$ frequency/time resource elements (REs) every 5 subframes (5 ms), while DM-RS requires $12$ REs per subframe just for $s=2$. Thus, DM-RS (downlink per-user pilot) is much more dense in time/frequency resource blocks.} than the common pilot, since the receiver performance is very sensitive to the time/frequency density of the former. As a result, the overhead of downlink per-user pilot is indeed a {more serious} limiting factor to the performance of realistic large-scale MIMO systems. Taking this into account, we let $N_1$ and $N_2$ be the frequency smoothness intervals for common pilot and downlink per-user pilot, respectively, with $N_1 > N_2$.

We consider the i.i.d. Rayleigh fading MIMO BC where $M=\alpha K$ with $\alpha\gg 1$, typical in large-scale MIMO.  
When $M<N_\text{LLN}$ and hence we need orthogonal resources for per-user pilots, the total number of degrees of freedom for communication is upper-bounded by 
\begin{align} \label{eq:PL-1}
   Q\left(T_c-\frac{Q}{N_1}-\frac{Q}{N_2}\right) 
\end{align}
where $\frac{Q}{N_1}$ and $\frac{Q}{N_2}$ are the number of channel uses needed to transmit common and per-user pilots, respectively.
It is easy to see the optimal number of scheduled users is given by $\min\{K,\frac{T_cN_1N_2}{2(N_1+ N_2)}\}$ since $M>K$. When $M$ is sufficiently larger than $N_\text{LLN}$, (\ref{eq:PL-1}) reduces to
$ Q\big(T_c-\frac{Q}{N_1}-1\big)$, 
yielding the optimal $Q$ of $\min\{K,\frac{(T_c-1)N_1}{2}\}.$ Fig. \ref{fig-6} shows how the performance regimes of large-scale MIMO systems are approximately shaped with these system parameters of interest, when $\alpha=10$ and $\frac{T_cN_1N_2}{2(N_1+ N_2)}< \frac{N_\text{LLN}}{10}< \frac{(T_c-1)N_1}{2}$. The shaded region represents the system performance loss due to the overhead of downlink per-user pilot, while the LLN region indicates that the saturated multiplexing gain starts increasing again, as long as $M\ge N_\text{LLN}$. Therefore, the per-user pilot overhead becomes a significant bottleneck of the performance of large-scale MIMO systems, unless both $M$ and $T_c$ are sufficiently large. If $T_c$ is small such that $ \frac{(T_c-1)N_1}{2}\le\frac{N_\text{LLN}}{\alpha}$, the favorable propagation condition due to LLN has no benefit in terms of the system multiplexing gain.

\begin{figure} 
\center\includegraphics[scale=1.2]{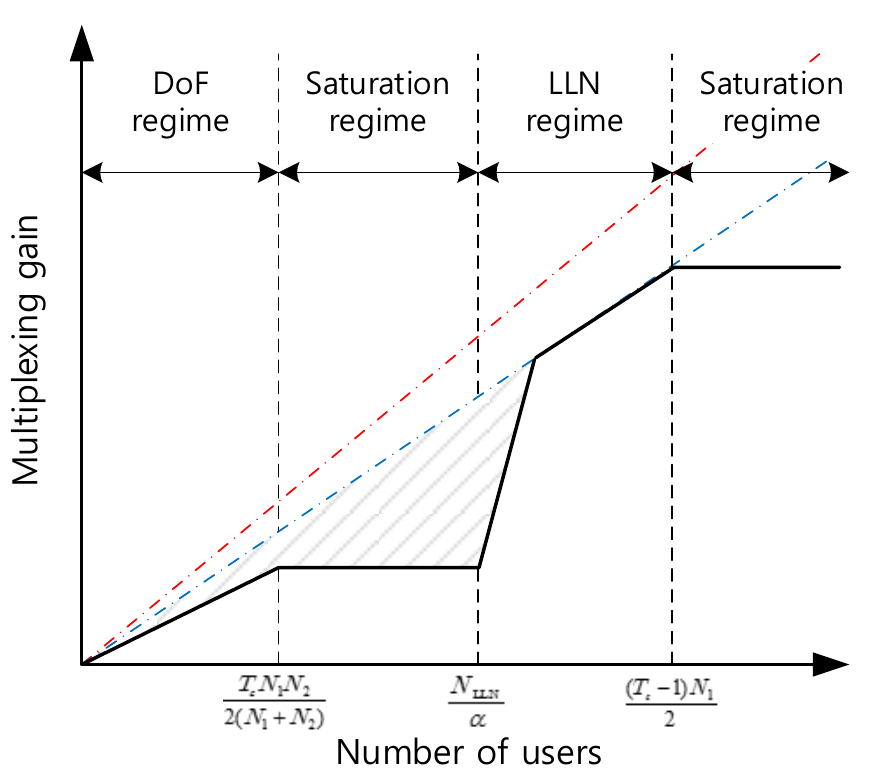} 
\caption{Performance regimes of conventional large-scale MIMO vs. the number of users ($K$) for $\frac{T_cN_1N_2}{2(N_1+ N_2)}< \frac{N_\text{LLN}}{\alpha}< \frac{(T_c-1)N_1}{2}$ and $M=\alpha K$.} \label{fig-6}
\end{figure}

Finally, it should be pointed out that 
if we consider the special case where all users have the same transmit correlation matrix $\pR$ with rank $r$, the threshold $N_\text{LLN}$ for the LLN region becomes larger by a factor of $M/r$ due to the effective channel dimension reduction in (\ref{eq:SM-5}). The resulting threshold may be too large to achieve the LLN region such that  $\frac{N_\text{LLN}M}{\alpha r} \ge \frac{(T_c-1)N_1}{2}$ and hence the degrees of freedom may be persistently saturated for $K\ge \frac{T_cN_1N_2}{2(N_1+ N_2)}$. If we cannot exploit transmit correlation diversity, it is evident that large-scale MIMO systems would suffer from severe degradation irrespectively of TDD or FDD, e.g., see \cite{Hoy13}, where all users have the common $\pR$ and the performance of large-scale MIMO with linear precoding/detection schemes was analyzed.

 



\section{Concluding Remarks}
\label{sec:CR}

In this paper, we have investigated several asymptotic capacity bounds of correlated fading MIMO BCs to understand the impact of transmit correlation on the capacity. In order to intuitively show the potential gains of transmit correlation diversity, we imposed the ideal unitary structure on channel covariances of users. Assuming perfect CSIT with no cost, we showed that transmit correlation diversity is not beneficial at all to the high-SNR capacity of Gaussian MIMO BCs in some regimes of system parameters like $M,K$, and $G$, while it helps the capacity in some other regimes.
Considering the cost for downlink training, we found that transmit correlation diversity is indeed very beneficial in that
multiplexing gain can continue growing as the number of antennas and the number of users increase, as long as transmit correlations at the BS are sufficiently high and well structured. 
Notice that the notion of transmit correlation diversity can be leveraged in various forms of MIMO wireless networks including MIMO MACs, multi-cell MU-MIMO systems, and wireless interference networks.

It was shown that the eigen-beamforming gain due to multiple pre-beamforming along long-term eigensapces is essential to achieve the capacity of correlated fading MIMO BCs. This provides an insight that a precoding scheme which can realize a large portion of such a gain is preferred to ZFBF for correlated fading channels particularly in the large $K$ regime. In order to validate this argument, our recent work in \cite{Nam14b} proposed a new limited feedback framework for large-scale MIMO systems. 

In MIMO wireless communications, there exist three most essential resources: time, frequency, and ``{small-scale}" space that depends on instantaneous channel realizations. Apart from these resources, we have identified a new type of resource, transmit correlation diversity (namely, ``{large-scale}" spatial resource), and provided an insight on how to use it and how it affects the system performance. The most remarkable result can be summarized as:
\emph{Exploiting transmit correlation may increase the multiplexing gain of MU-MIMO systems by a factor of the number of degrees of transmit correlation diversity.}

\section*{Acknowledgement} 
The author would like to thank Giuseppe Caire for his valuable comments to improve this work.

\appendices

\section{Useful Lemmas}  \label{app:lem}

We collect here some lemmas which are useful to prove theorems in this work.
 
\begin{lem}[\cite{Tul04}] \label{lem-1}
A central Wishart matrix $\Wm\Wm^\ct$ with $\Wm$ the $m\times n$ matrix, where $n\ge m$, satisfies 
\begin{align} \label{eq:ID-2}
   \mathbb{E} \big [\ln\left|  \Wm\Wm^\ct \right|\big ]  =  \sum_{\ell=0}^{m-1} \psi(n-\ell)
\end{align}
where 
\begin{align} \label{eq:ID-3}
   \psi(n) = -\gamma +\sum_{\ell=1}^{n-1}\frac{1}{\ell}
\end{align}
is the Euler's digamma function with $\gamma\approx 0.5572$ the Euler-Mascheroni constant. 
\end{lem}

The following lemma shows a useful asymptotic behavior of the central Wishart matrix.
 
\begin{lem} \label{lem-2}
For $m$ large with the ratio $\eta=\frac{n}{m}$ fixed, the central Wishart matrix $\Wm\Wm^\ct$ with $\Wm$ the $m\times n$ matrix, where $n\ge m$, shows the asymptotic behavior
\begin{align} \label{eq:ID-4}
   \frac{1}{m}\mathbb{E} \left [\ln\left|  \Wm\Wm^\ct \right|\right ]  =  \left(\eta-1\right)\ln \frac{\eta}{\eta-1} +\ln n-1 +O(m^{-1}).
\end{align}
\end{lem}

The proof of (\ref{eq:ID-4}) can be immediately given by applying 
\begin{align} \label{eq:ID-11}
  \frac{1}{k}\sum_{\ell=1}^{k} \psi(\ell)=\psi(k+1)-1
\end{align} 
and by using the fact that $\psi(k)$ behaves as
\begin{align} \label{eq:ID-10}
   \lim_{k\rightarrow \infty}\psi(k) =\ln k +O(k^{-1})
\end{align} 
due to $\lim_{k\rightarrow \infty} \sum_{n=1}^{k} \frac{1}{n} - \ln k = \gamma$.

The following lemma provides bounds on the determinant of the sum of two Hermitian matrices.

\begin{lem}[\cite{Fie71}] \label{lem-5}
Let $\Am$ and $\Bm$ be Hermitian matrices with eigenvalues $\lambda_1(\Am)\ge\lambda_2(\Am)\ge \cdots \ge \lambda_n(\Am)$ and $\lambda_1(\Bm)\ge\lambda_2(\Bm)\ge \cdots \ge \lambda_n(\Bm)$, respectively. Then
\begin{align} \label{UP-13}
  \min_\pi\prod_{i=1}^{n} \left(\lambda_i(\Am)+\lambda_{\pi(i)} (\Bm) \right ) \le \big|\Am+\Bm\big| \le   \max_\pi\prod_{i=1}^{n} \left(\lambda_i(\Am)+\lambda_{\pi(i)} (\Bm) \right )
  \end{align}
where $\pi$ denotes a permutation of indices $1,2,\cdots, n$.
In particular, if $\lambda_n(\Am)+\lambda_n(\Bm)\ge 0$, then
\begin{align} \label{UP-14}
  \prod_{i=1}^{n} \left(\lambda_i(\Am)+\lambda_{i} (\Bm) \right ) \le \big|\Am+\Bm\big| \le   \prod_{i=1}^{n} \left(\lambda_i(\Am)+\lambda_{n-i+1} (\Bm) \right ) .
  \end{align}
\end{lem}

We make use of the following simple bounds on $\trace(\ln \pA)$ based on Gaussian quadrature and related theory.

\begin{lem}[\cite{Bai96}] \label{lem-4}
Let $\pA$ be an $n\times n$ symmetric positive definite matrix, $\xi_1=\trace(\pA)$, $\xi_2=\trace(\pA^2)$, $\underline{\lambda}=\lambda_n(\Am)$, and $\overline{\lambda}=\lambda_1(\Am)$, then
\begin{align} \label{UP-30}
  \left[\ln \underline{\lambda} \;\ln \underline{t} \right] \left [\begin{matrix} \underline{\lambda} & \underline{t} \\ \underline{\lambda}^2 & \underline{t}^2 \end{matrix} \right ]^{-1} \left [\begin{matrix} \xi_1 \\ \xi_2  \end{matrix} \right ] 
  \le \trace(\ln \pA) \le  
  \left[\ln \overline{\lambda} \;\ln \overline{t} \right] \left [\begin{matrix} \overline{\lambda} & \overline{t} \\ \overline{\lambda}^2 & \overline{t}^2 \end{matrix} \right ]^{-1} \left [\begin{matrix} \xi_1 \\ \xi_2  \end{matrix} \right ] .
\end{align}
where $\underline{t} =\frac{\underline{\lambda}\xi_1 -\xi_2}{\underline{\lambda} n -\xi_1}$ and $\overline{t} =\frac{\overline{\lambda}\xi_1 -\xi_2}{\overline{\lambda} n -\xi_1}$.  
\end{lem}

\section{Proof of Theorem \ref{thm-2}}  
\label{app:proof-2}

We first prove the case of $r<K'$. Provided the unitary structure is available, the sum rate of the $g$th  dual MAC subchannel in (\ref{eq:MU-0}) can be rewritten as
\begin{align} \label{eq:IT-40}
   \log \left | \Id + \Lambdam_g^{1/2} \Wm_g \Sm_g \Wm_g^\mathsf{H} \Lambdam_g^{1/2} \right | &=  \log \left | \Lambdam_g^{-1} +  \Wm_g \Sm_g \Wm_g^\mathsf{H}  \right | +\log \left | \Lambdam_g \right |.  
\end{align}
By allowing the partial cooperation (i.e., the receiver cooperation within each group), based on the standard approach, the capacity region of the dual MAC subchannel is outer-bounded by that of the corresponding cooperative MIMO system. Given the perfect CSIT and at high SNR, the asymptotic optimal input $\Xm_g$ in the cooperative MIMO system is the uniform power allocation over $r$ eigenmodes of $\pW_g\pW_g^\ct$  with $\sum_g\trace(\Xm_g)\le P$, since the Wishart matrix $\pW_g\pW_g^\ct$ is well conditioned with high probability for all $g$. Here, the difference with our problem of interest is that the noise variances at $r$ effective antennas of the receiver in the $g$th dual MAC in the RHS of (\ref{eq:IT-40}) are scaled by $\lambda_{g,i}$. But this does not change the known result, since $\lambda_{g,i}$ are uniformly bounded by assumption in (\ref{eq:SM-1b}). Then, we have at high SNR (i.e., large $P$)
\begin{align} \label{eq:IT-35}
   \log \left | \Lambdam_g^{-1} +  \Wm_g \Sm_g \Wm_g^\mathsf{H}  \right |  &\le\log \left | \Lambdam_g^{-1} +  \Wm_g \Xm_g \Wm_g^\mathsf{H}  \right | \nonumber \\
   &\simeq \log \left | \Lambdam_g^{-1} +  \frac{P}{M}\Wm_g \Wm_g^\mathsf{H}  \right |  \nonumber \\
   &\simeq  \log \left |   \Wm_g \Wm_g^\mathsf{H}  \right | +r\log \frac{P}{M}
\end{align}
where $\simeq$ denotes the asymptotic equivalence (the difference between both sides vanishes as $P\rightarrow \infty$) and we used the fact that $\lambda_{g,i}^{-1}< \infty$ for all $i$ due to the uniform boundedness of $\lambda_{g,i}$.
As a consequence, when $r<K'$, the sum capacity is upper-bounded as 
\begin{align} 
  \mathcal{C}^\text{sum}(P) \le & \ M\log\frac{P}{M}+\sum_{g=1}^G \mathbb{E} \Big [\log\left|  \Wm_g \Wm_g^\ct \right|\Big ] +\sum_{g=1}^G\log|\Lambdam_g|+o(1) \label{eq:IT-37a} \\
  = & \ M\log\frac{P}{M}+\log e\sum_{g=1}^G \Big(K'\psi(K'+1)-(K'-r)\psi(K'-r+1)-r  \Big) +\sum_{g=1}^G\log|\Lambdam_g|+o(1) \nonumber \\
  = & \ M\log\frac{P}{M}+ rG\Bigg(-\gamma +\sum_{\ell=2}^{K'}\frac{1}{\ell}+\frac{K'-r}{r}\sum_{\ell=K'-r+1}^{K'}\frac{1}{\ell} \Bigg)\log e +\sum_{g=1}^G\log|\Lambdam_g|+o(1) \label{eq:IT-37}
\end{align}
where we used the well-known result of random matrix theory \cite{Tul04} in Lemma \ref{lem-1} of Appendix \ref{app:lem}, namely
\begin{align} \label{eq:IT-36}
   \mathbb{E} \Big [\ln\left|  \Wm_g \Wm_g^\ct \right|\Big ]  =  \sum_{\ell=0}^{r-1} \psi(K'-\ell) 
\end{align}
where $\Wm_g \Wm_g^\ct$ is almost surely nonsingular and $\psi(n)$ is defined in (\ref{eq:ID-3}). From (\ref{eq:ID-11}), the second equality in (\ref{eq:IT-37}) immediately follows.

The achievability of (\ref{eq:IT-13}) is given by simply letting the diagonal input matrix $\pS_g$ as $\pS_g=  \frac{P}{K} \pI_{K}$ for all $g$, which is in fact the optimal input covariance when only the channel distribution is accessible at the receiver in MIMO MAC with each user having the same power constraint \cite{Gol03}.  
The resulting $\Wm_g \Sm_g \Wm_g^\mathsf{H} = \frac{P}{K}{\Wm}_g {\Wm}_g^\mathsf{H}$ is also a Wishart matrix with $r$ degrees of freedom and hence  
\begin{align} \label{eq:IT-35b}
   \log \left | \Lambdam_g^{-1} +  \Wm_g \Sm_g \Wm_g^\mathsf{H}  \right |  &\ge  \log \left |  {\Wm}_g {\Wm}_g^\mathsf{H} \right | +r\log \frac{P}{K}+o(1)
\end{align}
Using (\ref{eq:ID-2}) in Lemma \ref{lem-1}, we have
\begin{align}  
  \mathcal{C}^\text{sum}(P) \ge  \ M\log\frac{P}{K}+ rG\Bigg(-\gamma +\sum_{\ell=2}^{K'}\frac{1}{\ell}+\frac{K'-r}{r}\sum_{\ell=K'-r+1}^{K'}\frac{1}{\ell} \Bigg)\log e +\sum_{g=1}^G\log|\Lambdam_g|+o(1) .
\end{align}
Then, we see that the achievable rate approaches the upper bound in (\ref{eq:IT-37}) within $c_{P,1}$ being $-M\log \frac{K'}{r} \le c_{P,1}\le 0$, which yields (\ref{eq:IT-13}).

Next, we consider the second case of $r \ge K'$. 
When the number of transmit antennas is greater than or equal to the total number of receive antennas in a MIMO BC, the sum capacity of its dual MAC is well known \cite{Cai03} to be equivalent at high SNR to that of the corresponding cooperative MIMO system. This also implies that, for $r \ge K'$, uniform power allocation across $K'$ eigenmodes in the $g$th dual MAC (\ref{eq:MU-0}) is asymptotically optimal, yielding $\pS_g=\frac{P}{K}\pI_{K'}$  for all $g$. As before, this well-known result holds true for our case due to the uniform boundedness of $\lambda_{g,i}$. Therefore, for sufficiently large $P$, we have the upper bound 
\begin{align} \label{eq:IT-28b}
      \log \left | \Lambdam_g^{-1} +  \Wm_g \Sm_g \Wm_g^\mathsf{H}  \right |  
   &\simeq \log \left | \Lambdam_g^{-1} +  \frac{P}{K}\Wm_g  \Wm_g^\mathsf{H}  \right |  \nonumber \\
   &\overset{(a)}{\le} \log \prod_{i=1}^{r} \left ( \lambda_{g,i}^{-1} +  \frac{P}{K}\lambda_i(\Wm_g  \Wm_g^\mathsf{H}) \right )  \nonumber \\
   &\overset{(b)}{=} \log \prod_{i=1}^{K'} \left ( \lambda_{g,i}^{-1} +  \frac{P}{K}\lambda_i(\Wm_g^\ct \Wm_g) \right ) +\log \prod_{i=K'+1}^{r} \lambda_{g,i}^{-1}  \nonumber \\
   &\simeq  \log \left |   \Wm_g^\ct \Wm_g \right | +K'\log \frac{P}{K} +\log \prod_{i=K'+1}^{r} \lambda_{g,i}^{-1} 
\end{align}
where $(a)$ follows from the upper bound of Lemma \ref{lem-5} in Appendix \ref{app:lem} and $(b)$ follows from the fact that the non-zero eigenvalues of $\Wm_g  \Wm_g^\mathsf{H}$ are the same as those of $\Wm_g^\ct \Wm_g$. Using the lower bound in Lemma \ref{lem-5}, we can similarly get 
\begin{align} \label{eq:IT-45}
      \log \left | \Lambdam_g^{-1} +  \Wm_g \Sm_g \Wm_g^\mathsf{H}  \right |  
   \ge  \log \left |   \Wm_g^\ct \Wm_g \right | +K'\log \frac{P}{K} +\log \prod_{i=K'+1}^{r} \lambda_{g,r-i+1}^{-1}. 
\end{align}

Plugging (\ref{eq:IT-28b}) and (\ref{eq:IT-45}) into (\ref{eq:IT-40}), using the fact that for $r \ge K'$, the Wishart matrix $\Wm_g^\ct \Wm_g$ is almost surely nonsingular, and invoking (\ref{eq:ID-2}) again, we have 
\begin{align}  
  K &\log\frac{P}{K}+K\Bigg(-\gamma +\sum_{\ell=2}^{r}\frac{1}{\ell} +\frac{r-K'}{K'}\sum_{\ell=r-K'+1}^{r}\frac{1}{\ell}\Bigg)\log e+\sum_{g=1}^G\log \prod_{i=1}^{K'} \lambda_{g,i}+o(1)\nonumber \\
 &\le   \mathcal{C}^\text{sum}(P)   \le  \nonumber \\ &K \log\frac{P}{K}+K\Bigg(-\gamma +\sum_{\ell=2}^{r}\frac{1}{\ell} +\frac{r-K'}{K'}\sum_{\ell=r-K'+1}^{r}\frac{1}{\ell} \Bigg)\log e+\sum_{g=1}^G\log \prod_{i=1}^{K'} \lambda_{g,r-i+1}+o(1).
\end{align}
With $c_{P,2}$ being $\sum_{g=1}^G\sum_{i=1}^{K'}\log\frac{\lambda_{g,r-i+1}}{\lambda_{g,i}} \le c_{P,2}\le 0$, we have (\ref{eq:IT-14}). This completes the proof.

Beside the above proof, an alternative expression of (\ref{eq:IT-14}) can be found as follows. 
For sufficiently large $P$, we have
\begin{align} \label{eq:IT-44}
   \log \left | \Id + \Lambdam_g^{1/2} \Wm_g \Sm_g \Wm_g^\mathsf{H} \Lambdam_g^{1/2} \right | &\simeq  \log \left | \pI+\frac{P}{M} \Lambdam_g^{1/2} \Wm_g  \Wm_g^\mathsf{H} \Lambdam_g^{1/2} \right | \nonumber \\
   &= \log \left |\pI+\frac{P}{M} \Wm_g^\mathsf{H}\Lambdam_g \Wm_g    \right | \nonumber \\
   &\simeq \log \left | \Wm_g^\mathsf{H}\Lambdam_g \Wm_g    \right |  + r\log \frac{P}{M}.  
\end{align}
Using Lemma 2 in \cite{Loz05} (See also \cite{Shi03}), we can get
\begin{align}  
  \mathbb{E} \Big [\log\left| \Wm_g^\mathsf{H}\Lambdam_g \Wm_g\right|\Big ] =\log e\sum_{g=1}^G \frac{| \boldsymbol{\Upsilon}_g|}{|\boldsymbol{\Omega}_g|}\sum_{k=1}^{K'}|\boldsymbol{\Psi}_{g,k}|  
\end{align}
where $\boldsymbol{\Psi}_{g,k}$ is an $K'\times K'$ matrix whose $(i,j)$ element is 
\begin{align} \label{eq:IT-16}
  \big(\boldsymbol{\Psi}_{g,k}\big)_{i,j}=\nu_{r-K'+i}\lambda_{g,r-K'+i}^{r-K'-1+j} -\sum_{d=1,q=1}^{r-K'}\nu_q\big(\boldsymbol{\Upsilon}_g^{-1}\big)_{d,q} \lambda_{g,r-K'+i}^{d-1}\lambda_{q}^{g,r-K'-1+j} 
\end{align}
where $\nu_q = \psi(\ell)+\ln\lambda_q$ for $\ell= k$; otherwise, $\nu_q =1$, $\boldsymbol{\Omega}_g$ is the Vandermonde matrix
\begin{align} \label{eq:IT-17}
  \boldsymbol{\Omega}_g =\left [\begin{matrix} 1 & \lambda_{g,1} & \cdots & \lambda_{g,1}^{r-1} \\ 1 & \lambda_{g,2} & \cdots & \lambda_{g,2}^{r-1} \\  \vdots & \vdots & \ddots & \vdots \\ 1 & \lambda_{g,r} & \cdots & \lambda_{g,r}^{r-1}\end{matrix} \right ]
\end{align}
and $\boldsymbol{\Upsilon}_g$ is the $(r-K')\times(r-K')$ principle submatrix of $\boldsymbol{\Omega}_g$.
This yields 
\begin{align}  \label{eq:IT-43}
  \mathcal{C}^\text{sum}(P)= M\log\frac{P}{M}+\log e\sum_{g=1}^G \frac{| \boldsymbol{\Upsilon}_g|}{|\boldsymbol{\Omega}_g|}\sum_{k=1}^{K'}|\boldsymbol{\Psi}_{g,k}| +o(1).
\end{align}

\section{Proof of Theorem \ref{thm-8}}  
\label{app:proof-8}

The proof begins with the dual MAC in  (\ref{eq:MU-0}) divided by $M$
\begin{align} \label{eq:ID-7}
   \frac{1}{M}\sum_{g=1}^G\log \left | \Id + \Lambdam_g^{1/2} \Wm_g \Sm_g \Wm_g^\mathsf{H} \Lambdam_g^{1/2} \right | 
   &= \frac{1}{M} \sum_{g=1}^G\log \left | \Lambdam_g^{-1} +  \Wm_g \Sm_g \Wm_g^\mathsf{H}  \right | +\frac{1}{M}\sum_{g=1}^G \log \left | \Lambdam_g \right |
\end{align}
where the equality is given by (\ref{eq:IT-40}) and the assumptions. For $\mu < 1$ (i.e., $r<K'$) at high SNR ($P$), taking expectation on the first term in the right-hand side (RHS) of (\ref{eq:ID-7}), we have the upper bound
\begin{align} \label{eq:ID-8}
   \frac{1}{M}\mathbb{E}\left [\sum_{g=1}^G\log \left | \Lambdam_g^{-1} +  \Wm_g \Sm_g \Wm_g^\mathsf{H}  \right | \right ] &\overset{(a)}{\le}  \frac{1}{r}\;\mathbb{E} \Big [\log \left |   \Wm_g \Wm_g^\mathsf{H}  \right | \Big ] +\log \frac{P}{M}  \nonumber \\
   &\overset{(b)}{=} \log e \left\{(\mu^{-1}-1)\ln\frac{\mu^{-1}}{\mu^{-1}-1} +\ln K'-1+O(r^{-1}) \right\} +\log \frac{P}{M}\nonumber \\
   &= \log \frac{P}{e\mu G}+\Big(\frac{1-\mu}{\mu}\Big)\log\frac{1}{1-\mu} +O(r^{-1})
\end{align}
where $(a)$ follows from (\ref{eq:IT-35}) and $(b)$ follows from (\ref{eq:ID-4}) in Lemma \ref{lem-1}. From (\ref{eq:IT-35b}) and (\ref{eq:ID-4}), we also get the lower bound
\begin{align} \label{eq:ID-9}
   \frac{1}{M}\mathbb{E}\left [\sum_{g=1}^G\log \left | \Lambdam_g^{-1} +  \Wm_g \Sm_g \Wm_g^\mathsf{H}  \right | \right ] 
   &\ge \log \frac{P}{e G}+\Big(\frac{1-\mu}{\mu}\Big)\log\frac{1}{1-\mu} +O(r^{-1}).
\end{align}

The second term in the RHS of (\ref{eq:ID-7}) can be bounded by using 
\begin{align} \label{eq:ID-20}
   \log\lambda_\text{min} \le \frac{1}{M}\sum_{g=1}^G\log \left | \Lambdam_g \right | \le \frac{1}{M}\sum_{g=1}^G\log \left(\frac{\trace(\Lambdam_g)}{r}\right)^r =\log G
\end{align}
where $\lambda_\text{min}=\min_g \lambda_{g,r}$ and we used $\trace(\pR_g)=\trace(\Lambdam_g)=M$ and the geometric-arithmetic mean inequality $|\pA| \le \big(\frac{\trace(\pA)}{r}\big)^r$ with $r$ the rank of an $n\times n$ matrix $\pA$. 
Combining (\ref{eq:ID-8}) -- (\ref{eq:ID-28}), we obtain (\ref{eq:ID-5}).  


For $\mu \ge 1$ (i.e., $r\ge K'$) and large $r$, similar to the above steps with 
\begin{align} \label{eq:ID-21}
    \log\lambda_\text{min} &\le \frac{1}{K'} \sum_{g=1}^G\log\prod_{i=1}^{K'} \lambda_{g,r-i+1} \nonumber \\ &\le \frac{1}{K'} \sum_{g=1}^G\log\prod_{i=1}^{K'} \lambda_{g,i}\nonumber \\ & \le \frac{1}{K'}\log \left(\frac{\sum_{i=1}^{K'}\lambda_{g,i}}{K'}\right)^{K'}\nonumber \\ &\le \log \left(\frac{M}{K'}\right)= \log \mu G 
\end{align}
we can obtain (\ref{eq:ID-6}) by using (\ref{eq:IT-28b}), (\ref{eq:IT-45}) and (\ref{eq:ID-4}). 
The remaining detail is omitted for the sake of the compactness of this paper. 

\section{Achievability of (\ref{eq:TC-3})}  
\label{app:proof-3}

The achievability proof of (\ref{eq:TC-3}) begins with (\ref{eq:IT-40}), Corollary 1 in \cite{Nam13a}, and the uniform power allocation over groups such that $\pS_g=\frac{P}{M}\pI_{r}$, yielding
\begin{align}  \label{eq:TC-3b}
  \sum_{g=1}^G r\log\log K' +M\log\frac{P}{M}+\sum_{g=1}^G\log|\Lambdam_{g}|+o(1).
\end{align}
Compared to $M\log\log K$ in the i.i.d. Rayleigh fading case, the multiuser diversity gain reduces to $\sum_{g=1}^G r\log\log K'$. To show that this diversity gain reduction vanishes for sufficiently large $K'$, we use the logarithmic identity
\begin{align} \label{eq:IT-31}
  \log_c(a\pm b)=\log_c a+\log_c\left(1\pm\frac{b}{a}\right)
\end{align}
where $a$ and $b$ are nonnegative. Then, we get 
$$\sum_{g=1}^G r\log\log K'=M\log\log K+o(1)$$
for large $K'$. This proves the achievability.

\bibliographystyle{IEEEtran}
\bibliography{transmit_correlation_diversity_final_v1}

\end{document}

%% file: macros.tex
\long\def\comment#1{}

\newfont{\bbb}{msbm10 scaled 700}

\newfont{\bbc}{msbm10 scaled 1100}

\newcommand{\wv}{{\pmb w}}

\newcommand{\Am}{{\pmb A}}
\newcommand{\Bm}{{\pmb B}}

\newcommand{\Hm}{{\pmb H}}
\newcommand{\Id}{{\pmb I}}

\newcommand{\Sm}{{\pmb S}}

\newcommand{\Vm}{{\pmb V}}
\newcommand{\Wm}{{\pmb W}}
\newcommand{\Xm}{{\pmb X}}

\newcommand{\Cc}{{\cal C}}

\newcommand{\Lambdam}{\hbox{\boldmath$\Lambda$}}

\newcommand{\Sigmam}{\hbox{\boldmath$\Sigma$}}

\newcommand{\diag}{{\hbox{diag}}}
\renewcommand{\det}{{\hbox{det}}}
\newcommand{\trace}{{\hbox{tr}}}
